\documentclass[prb,twocolumn,preprintnumbers,amsmath,amssymb,floatfix,superscriptaddress]{revtex4-1}

\usepackage{graphicx,color,hyperref}
\usepackage[caption=false]{subfig}
\begin{document}

\title{Diffusion quantum Monte Carlo and $GW$ study of the electronic
  properties of monolayer and bulk hexagonal boron nitride}

\author{R.\ J.\ Hunt}

\affiliation{Department of Physics, Lancaster University, Lancaster
  LA1 4YB, United Kingdom}

\author{B.\ Monserrat}

\affiliation{Cavendish Laboratory, University of Cambridge,
  J.\ J.\ Thomson Avenue, Cambridge CB3 0HE, United Kingdom}

\author{V.\ Z\'{o}lyomi}

\affiliation{Hartree Centre, STFC Daresbury Laboratory, Daresbury WA4
  4AD, United Kingdom}

 \author{N.\ D.\ Drummond}

\affiliation{Department of Physics, Lancaster University, Lancaster
  LA1 4YB, United Kingdom}

\date{\today}

\begin{abstract} We report diffusion quantum Monte Carlo (DMC) and many-body
  $GW$ calculations of the electronic band gaps of monolayer and bulk
  hexagonal boron nitride (hBN)\@.  We find the monolayer band gap to
  be indirect.  $GW$ predicts much smaller quasiparticle gaps at both
  the single-shot $G_0W_0$ and the partially self-consistent $GW_0$
  levels. In contrast, solving the Bethe-Salpeter equation on top of
  the $GW_0$ calculation yields an exciton binding energy for the
  direct exciton at the $K$ point in close agreement with the DMC
  value.  Vibrational renormalization of the electronic band gap is
  found to be significant in both the monolayer and the bulk.  Taking
  vibrational effects into account, DMC overestimates the band gap of
  bulk hBN, while $GW$ theory underestimates it.
\end{abstract}

\maketitle


\section{Introduction\label{sec:intro}}

Two-dimensional (2D) materials have provided an exciting new frontier
for experimental and theoretical nanoscience in the fifteen years
since the first isolation of atomically thin layers of graphene by
mechanical exfoliation from graphite \cite{Novoselov_2004,Geim_2007}.
In addition to graphene and its derivatives, the last few years have
witnessed growing interest in semiconducting 2D materials such as
transition-metal dichalcogenides
\cite{Wang_2012,Mak_2010,Splendiani_2010} and phosphorene
\cite{Li_2014,Koenig_2014,Liu_2014}.  A recent trend has been the
study of stacked heterostructures of 2D materials \cite{Geim_2013}.
Heterostructures involving graphene and hexagonal boron nitride (hBN)
have received particular attention, because monolayer hBN is an
insulating, atomically thin 2D material with a similar lattice
constant to graphene and is therefore the ideal substrate for
graphene-based electronics
\cite{Dean_2010,Xue_2011,Bresnehan_2012,Lee_2012,Liu_2013}.  Monolayer
or few-layer hBN is potentially an important component in novel
electronic devices based on 2D materials, such as vertical tunneling
diodes \cite{Britnell_2012,Britnell_2013} and supercapacitors
\cite{Shi_2014}. In addition, due to the slight lattice mismatch,
graphene placed on hBN exhibits a moir\'{e} pattern with a period of
up to 14 nm \cite{Yankowitz_2012}, and the resulting superlattices
allow the experimental observation of exotic phenomena such as the
formation of Hofstadter's butterfly \cite{Hofstadter_1976} features in
the band structure in the presence of a magnetic field
\cite{Ponomarenko_2013,Dean_2013}. Despite the importance of 2D hBN in
current or proposed graphene-based electronics research, the
properties of monolayers of hBN are not currently well characterized
due to the experimental challenge of isolating and studying
monolayers.  In this paper, we use advanced theoretical
electronic-structure methods to provide basic information about the
size and nature of the electronic band gap of monolayers of hBN\@.  We
find that the gap of hBN monolayers is in principle indirect (so that
optical transitions involve the absorption or emission of phonons),
and that the quasiparticle gap is considerably enhanced relative to
the bulk.  However, the conduction band around its minimum at the
$\Gamma$ point is a free-electron-gas-like state that is only weakly
bound to the hBN monolayer and has a relatively small spatial overlap
with the valence states \cite{Blase_1995}; hence the dipole matrix
element for an optical transition from the valence-band maximum to the
conduction-band minimum is inevitably small.  Furthermore, the precise
energy of a state that extends outside the layer will be strongly
affected by the environment in which the layer finds itself
(substrate, encapsulation, etc.).  Hence we expect inverse
photoemission measurements to show the energy of the conduction band
at $\Gamma$ to depend strongly on the environment.  Likewise, the
effective height of the energy barrier presented by an hBN monolayer
in a vertical-tunneling experiment will depend sensitively on the
environment of the layer.

Bulk hBN (also known as white graphite) consists of layers of boron
and nitrogen atoms occupying the $A$ and $B$ hexagonal sublattice
sites of a 2D honeycomb lattice.  These layers are weakly bound
together by van der Waals interactions, resulting in both the
lubricating properties of hBN and the possibility of isolating
monolayers by mechanical exfoliation.  Bulk hBN adopts an $AA^\prime$
stacking arrangement in which each boron atom (with a positive partial
charge) has a nitrogen atom (with a negative partial charge)
vertically above it and \textit{vice versa}.  Whereas pristine
graphene is a gapless semiconductor, monolayer hBN is an insulator due
to the lack of sublattice symmetry.  Bulk hBN is semiconducting, with
experimental estimates of the band gap ranging from 5.2(2)--7.1(1) eV
\cite{Hoffmann_1984,Carpenter_1982,Watanabe_2004,Shi_2010,Cassabois_2016}.
Watanabe \textit{et al.}\ find the quasiparticle band gap to be direct
and of value 5.971 eV in a single-crystal sample \cite{Watanabe_2004}.
More recent experimental work by Cassabois \textit{et al.}\ has
indicated that bulk hBN is in fact an indirect semiconductor with a
quasiparticle band gap of 6.08 eV \cite{Cassabois_2016}.  The
experimental work of Cassabois \textit{et al.}, together with
subsequent theoretical works \cite{Cannuccia_2019,Paleari_2019}, have
elucidated the role of vibrational effects in phonon-assisted indirect
optical transitions in bulk hBN\@.  Many-body $GW$ calculations also
indicate that bulk hBN is an indirect-gap semiconductor, with a
fundamental gap of 5.95--6.04 eV between the valence-band maximum
(which is near the $K$ point, on the $\Gamma \rightarrow K$ line) and
the conduction-band minimum at $M$
\cite{Cappellini_1996,Arnaud_2006,Wirtz_2006}.

One of the many reasons for the high levels of interest in 2D
materials is that the electronic properties of monolayers often differ
significantly from those of the bulk layered material.  Density
functional theory (DFT) within the local density approximation (LDA)
predicts the indirect band gap of monolayer hBN to be 4.6 eV
\cite{Blase_1995}, and the $GW_0$ shift in the quasiparticle band gap
is about 3.6 eV \cite{Wirtz_2006}, giving a gap of 8.2 eV for the
monolayer.  Clearly the gap is considerably enhanced on going from
bulk hBN to a monolayer.  Bulk hBN is believed to exhibit a large
exciton binding energy, with values of 0.7--1.2 eV
\cite{Wirtz_2006,Arnaud_2006,Cunningham_2018} predicted by
$GW$-Bethe-Salpeter-equation ($GW$-BSE) calculations.  On the other
hand, experimental measurements find the exciton binding energy to be
only 0.13--0.15 eV \cite{Watanabe_2004,Cassabois_2016}, although there
are questions over the interpretation of these experimental results
\cite{Paleari_2019}. Exciton binding is further enhanced in a
free-standing monolayer due to the reduction in screening. Indeed,
$GW$-BSE calculations find that the exciton binding energy increases
to 2.1 eV in the monolayer \cite{Wirtz_2006}.

Isolating monolayer hBN by exfoliation from bulk hBN has proved
challenging, although Elias \textit{et al.}\ have recently succeeded
in growing atomically thin samples of hBN on graphite substrates
\cite{Elias_2019}. Their reflectance and photoluminescence
measurements indicate a direct gap of 6.1 eV for hBN on graphite.
However, the electronic properties of an isolated hBN monolayer (i.e.,
a freely suspended sample) are at present only accessible through
theoretical calculations. Unfortunately, DFT systematically
underestimates electronic band gaps and even many-body $GW$ methods
\cite{Hedin_1965} suffer from limitations, as exemplified by the
disagreement between the self-consistent and non-self-consistent
variants of the method when applied to hBN \cite{Wirtz_2006}. We have
therefore made use of quantum Monte Carlo (QMC) methods
\cite{Ceperley_1980,Foulkes_2001} to study many-body effects in the
band gap.  We have calculated the electronic band gaps for excitations
from the valence band at the $K$ point of the hexagonal Brillouin zone
($K_{\rm v}$) to the conduction band at the $\Gamma$ and $K$ points
($\Gamma_{\rm c}$ and $K_{\rm c}$) of monolayer hBN\@.  In our DFT and
$GW$ calculations, and our QMC calculations for bulk hBN, we have also
considered the conduction band at the $M$ point ($M_{\rm c}$).
Furthermore, we have investigated the effects of the vibrational
renormalization of the electronic structure at the DFT level.

We have made use of two QMC methods: variational Monte Carlo (VMC) and
diffusion Monte Carlo (DMC) \cite{Foulkes_2001}. In VMC, Monte Carlo
integration is used to evaluate quantum mechanical expectation values
with respect to trial wave-function forms of arbitrary complexity.
Free parameters in the trial wave functions are optimized by a
variational approach.  DMC involves simulating drifting, diffusion,
and birth/death processes governed by the Schr\"{o}dinger equation in
imaginary time to project out the ground-state component of a trial
wave function \cite{Ceperley_1980}. The fixed-node
approximation \cite{Anderson_1976} is used to maintain fermionic
antisymmetry.  All our QMC calculations were performed using the
\textsc{casino} code \cite{casino}.

QMC methods have only recently been applied to calculate the energy
gaps of 2D materials \cite{Hunt_2018,Frank_2018}. A major challenge
is the need to extrapolate the QMC band gaps to the thermodynamic
limit of large system size, because the computational expense of the
method necessitates the use of relatively small simulation supercells
subject to periodic boundary conditions \cite{Hunt_2018}. In this
article we investigate finite-size effects in the band gap of hBN\@.

The rest of this article is arranged as follows.  In
Sec.\ \ref{sec:methodology} we describe our DFT, $GW$, and QMC
methodologies.  We present our results in Sec.\ \ref{sec:results}.
Finally we draw our conclusions in Sec.\ \ref{sec:conclusions}.  We
use Hartree atomic units (a.u.)\ throughout, in which $\hbar=m_{\rm
  e}=|e|=4\pi\epsilon_0=1$, except where other units are given
explicitly.

\section{Computational methodology\label{sec:methodology}}

\subsection{DFT}

\subsubsection{Geometry optimization, lattice dynamics, and
band-structure calculations \label{sec:dft_methodology}}

We performed our DFT calculations using the LDA, the
Perdew-Burke-Ernzerhof (PBE) generalized-gradient-approximation
exchange-correlation functional \cite{Perdew_1996}, and the
Heyd-Scuseria-Ernzerhof (HSE06) hybrid
functional \cite{Heyd_2003,Krukau_2006}. We used the
\textsc{castep} \cite{castep} and \textsc{vasp} \cite{vasp}
plane-wave-basis DFT codes.  Our DFT-LDA and DFT-PBE relaxations of
the lattice parameter used an artificial periodicity of 21.17 {\AA} in
the out-of-plane direction, a $53 \times 53$ Monkhorst-Pack ${\bf
  k}$-point grid, ultrasoft pseudopotentials, and a plane-wave cutoff
of 680 eV\@.  The same parameters were used in our calculations of the
electronic band structure. Our phonon calculations used density
functional perturbation theory \cite{castep_dfpt}, norm-conserving DFT
pseudopotentials, a plane-wave cutoff energy of 1361 eV, an artificial
periodicity of 26.46 {\AA}, and a $53 \times 53$ Monkhorst-Pack ${\bf
  k}$-point grid for both the electronic orbitals and the vibrational
normal modes.  In our DFT-HSE06 calculations of the lattice parameter
and band structure we used an artificial periodicity of 15.875 {\AA}
in the out-of-plane direction, an $11 \times 11$ Monkhorst-Pack ${\bf
  k}$-point grid, norm-conserving DFT pseudopotentials, and a
plane-wave cutoff of 816 eV\@.

\subsubsection{QMC trial wave function generation\label{sec:dft_wf_gen}}

The DFT calculations performed to generate trial wave functions for
our QMC calculations used Dirac-Fock
pseudopotentials \cite{Trail_2005a,Trail_2005b}, a plane-wave cutoff
energy of 2721 eV, and, in the monolayer case, an artificial periodicity of
18.52 {\AA} (apart from the $3\times 3$ supercell, where the plane-wave cutoff
energy and artificial periodicity were 2177 eV and 13.35 {\AA}, respectively).

We found that replacing PBE ultrasoft pseudopotentials with
Trail-Needs Dirac-Fock pseudopotentials changes the monolayer $K_{\rm
  v} \rightarrow \Gamma_{\rm c}$ and $K_{\rm v} \rightarrow K_{\rm c}$
DFT-PBE gaps from $4.69$ to $4.71$ eV and from $4.67$ to $4.79$ eV,
respectively.  In these calculations the lattice parameter is fixed at
the DFT-PBE value, $a=2.512$ {\AA}, obtained with the PBE ultrasoft
pseudopotentials.  This suggests that the choice of pseudopotential
introduces an uncertainty of around $0.1$ eV into our QMC gap
estimates.

\subsection{$GW$(-BSE) calculations}

In the $GW$ approximation, many-body interactions are taken into
account in a quasiparticle picture in which the screened Coulomb
interaction $W$ between particles is included in the self-energy to
first order. Varying levels of approximation are possible: the
so-called single-shot $G_0 W_0$ approach calculates the Green's
function $G$ and the dielectric screening in the Coulomb interaction
$W$ from DFT wave functions, while the partially and fully
self-consistent $GW_0$ and $GW$ methods iterate one or both of these
quantities until self-consistency is achieved. Excitonic effects in
the optical absorption can be taken into account by solving the
Bethe-Salpeter equation (BSE) following the $GW$ calculations.  We
performed $G_0 W_0$(-BSE) and $GW_0$(-BSE) calculations for monolayer
hBN, and for test purposes also $G_0 W_0$ and $GW_0$ calculations for
bulk hBN\@.

In our $GW$ calculations we used the \textsc{vasp} \cite{vasp}
plane-wave-basis code for bulk hBN\@.  The HSE06
functional \cite{Heyd_2003,Krukau_2006} was used to calculate the
orbitals and their derivatives as input for the single-shot $G_0W_0$
calculations \cite{Shishkin_2006}. Convergence of the $G_0W_0$
calculation with respect to its principal convergence parameters was
achieved using a $12 \times 12 \times 12$ Monkhorst-Pack ${\bf
  k}$-point grid, with 24 electronic bands taken into account, and a
plane-wave cutoff energy of 400 eV\@. These parameters converge the
band gap of bulk hBN to within 0.1 eV\@. We used the same parameter
set to compute the partially self-consistent $GW_0$ band gap. The
results of the bulk calculations are discussed in
Sec.\ \ref{sec:dmc_gap_results}.

For the monolayer $GW$ calculations we used the \textsc{BerkeleyGW}
code \cite{berkeleygw} in order to be able to treat a much larger
number of empty bands. In the bulk, the $G_0W_0$ gap changes by less
than 50 meV when the number of electronic bands is increased from 24
to 48. In contrast, the monolayer requires 1200 bands to be taken into
account for the same level of convergence; otherwise the dielectric
function is too inaccurate to predict reliable self-energy
corrections. The ${\bf k}$-point grid for the monolayer calculations
was set to $24 \times 24 \times 1$, while the plane-wave cutoff during
the many-body calculations was set to 408.17 eV (30 Ry). In these
calculations, the DFT wave functions were calculated using the PBE
functional.  For the monolayer, the optical absorption coefficient was
also calculated at both the single-shot and the $GW_0$ level by
solving the Bethe-Salpeter equation. In both cases we took 6 empty and
4 occupied bands into account. Truncation of the Coulomb interaction was
applied in the monolayer calculations.

\subsection{QMC calculations}

\subsubsection{Evaluating quasiparticle and excitonic gaps\label{sec:eval_gaps}}

To calculate an excitation energy using DMC we exploit the fixed-node
approximation \cite{Anderson_1976} and evaluate the difference of the
total energies obtained using trial wave functions corresponding to
the ground state and the particular excited state of
interest \cite{Mitas_1994,Williamson_1998,Towler_2000}. For each
excited state an appropriate wave function can be constructed by
choosing the occupancies of the orbitals in the Slater determinants (see
Sec.\ \ref{sec:trial_wfs}).  The DMC energy of an excited state is
exact if the nodal surface of the trial wave function is exact, as is
the case for the ground state, although the DMC energy is only
guaranteed to be an upper bound on the energy for certain excited
states \cite{Foulkes_1999}.

The quasiparticle bands at a particular point may be evaluated as
${\cal E}_i({\bf k})=E^+({\bf k},i)-E^{\rm GS}$ for unoccupied states,
where $E^+({\bf k},i)$ is the total energy when an electron is added
to the system and occupies band $i$ at wavevector ${\bf k}$ and
$E^{\rm GS}$ is the ground-state total energy.  For occupied states we
evaluate ${\cal E}_i({\bf k})=E^{\rm GS}-E^-({\bf k},i)$, where
$E^-({\bf k},i)$ is the total energy when an electron is removed from
band $i$ at wavevector ${\bf k}$.  The quasiparticle band gap
$\Delta_{\rm qp}$ is the difference of the energy bands at the
conduction-band minimum (CBM) and valence-band maximum (VBM):
\begin{equation} \Delta_{\rm qp}={\cal E}_{\rm CBM}-{\cal E}_{\rm VBM}=E^+_{\rm
    CBM}+E^-_{\rm VBM}-2E^{\rm GS}. \end{equation}

The excitonic gap $\Delta_{\rm ex}$ is defined as the difference in
energy when an electron is promoted from the VBM to the CBM\@:
\begin{equation} \Delta_{\rm ex}=E^{\rm pr}_{{\rm VBM}\rightarrow {\rm
      CBM}}-E^{\rm GS}, \end{equation} where $E^{\rm pr}_{{\rm
    VBM}\rightarrow {\rm CBM}}$ is the total energy evaluated with a
trial wave function in which an electron has been promoted from the
VBM to the CBM\@.  In Sec.\ \ref{sec:spin_states} we investigate
whether it is important to construct appropriate wave functions for
excitonic spin singlets or triplets when calculating gaps.

\subsubsection{Trial wave functions \label{sec:trial_wfs}}

We used Slater-Jastrow (SJ) trial wave functions $\Psi=\exp(J)
S^\uparrow S^\downarrow$ in our QMC calculations.  The Slater
determinants for up- and down-spin electrons $S^\uparrow$ and
$S^\downarrow$ contained Kohn-Sham orbitals generated using the
\textsc{castep} plane-wave-basis code \cite{castep} in a
three-dimensionally periodic cell, as described in
Sec.\ \ref{sec:dft_wf_gen}.  However, the orbitals were re-represented
in a localized ``blip'' B-spline basis \cite{Alfe_2004} to improve the
scaling of the QMC calculations with system size and to allow us, in
the monolayer case, to discard the artificial periodicity in the
out-of-plane direction.  The Jastrow factor $\exp(J)$ is a positive,
symmetric, explicit function of interparticle distances. We used the
Jastrow form described in Ref.\ \onlinecite{Drummond_2004}, in which
the Jastrow exponent $J$ consists of short-range, isotropic
electron-electron, electron-nucleus, and electron-electron-nucleus
terms, which are polynomials in the interparticle distances, as well
as long-range electron-electron terms expanded in plane waves.  The
free parameters in our Jastrow factors were optimized by unreweighted
variance minimization \cite{Umrigar_1988a,Drummond_2005}.

Within Hartree-Fock theory, band gaps are significantly overestimated
due to the tendency to over-localize electronic states in a theory
that does not allow correlation to keep electrons apart.  DMC
retrieves a large but finite fraction of the correlation energy.
Assuming the fraction of correlation energy retrieved in the ground
state is similar to or greater than the fraction retrieved in an
excited state, we expect the DMC gaps to be upper bounds on the true
gaps.  If we increase the fraction of correlation energy retrieved,
e.g., by including a backflow
transformation \cite{Kwon_1993,Lopez_2006}, then (if anything) we
expect to see a decrease in the band gap.

We performed some test calculations with Slater-Jastrow-backflow (SJB)
trial wave functions \cite{Kwon_1993,Lopez_2006}. In a backflow wave
function the orbitals in the Slater determinant are evaluated not at
the actual electron positions, but at quasiparticle positions that are
functions of all the particle coordinates.  The backflow function,
which describes the offset of the quasiparticle coordinates relative
to the actual coordinates, contains free parameters to be determined
by an optimization method.  The Jastrow factor and backflow functions
were optimized by VMC energy minimization \cite{Umrigar_2007}. As shown
in Table \ref{table:dmc_energies}, backflow lowers the DMC total
energies significantly.  However the amount by which backflow reduces
the quasiparticle and excitonic gaps is small: about $0.10(3)$ eV on
average.  We investigated the reoptimization of backflow functions in
the supercells in which an electron has been added or removed, finding that
reoptimization raises the gap slightly.  This is perhaps indicative of static
correlation (multireference character) effects in the nodal surface
that are not addressed by the use of backflow.  Since QMC simulations
with backflow are significantly more expensive, and finite-size
effects are a potentially dominant source of error in our work, we did
not use backflow in our production calculations.

\begin{table*}[!htbp]
\begin{center}
\caption{DMC ground-state (GS) energy per primitive cell,
  quasiparticle band gap, and excitonic band gap of monolayer hBN
  in a supercell consisting of $3 \times 3$ primitive cells, as
  obtained using different trial wave functions and time steps.  The
  ${\bf k}$-vector grid includes both $\Gamma$ and $K$. Where the time
  step is 0, the reported results have been obtained by linear
  extrapolation to zero time step. The fact that the excitonic gap is
  higher than the quasiparticle gap is a manifestation of finite-size
  error in the uncorrected gaps, as explained in
  Sec.\ \ref{sec:finite_size}. \label{table:dmc_energies}}

\begin{tabular}{lr@{}lr@{}lr@{}lr@{}lr@{}lr@{}l}
\hline \hline

& \multicolumn{2}{c}{Time step} & \multicolumn{2}{c}{GS energy $E^{\rm
    GS}$} & \multicolumn{4}{c}{Quasiparticle gap $\Delta_{\rm qp}$ (eV)} &
\multicolumn{4}{c}{Excitonic gap $\Delta_{\rm ex}$ (eV)} \\

\raisebox{1.5ex}[0pt]{Wave fn.} & \multicolumn{2}{c}{(a.u.)} &
\multicolumn{2}{c}{(eV/p.\ cell)} & \multicolumn{2}{c}{~~~~$K_{\rm
    v}\rightarrow \Gamma_{\rm c}$} & \multicolumn{2}{c}{$K_{\rm
    v}\rightarrow K_{\rm c}$} & \multicolumn{2}{c}{$K_{\rm
    v}\rightarrow \Gamma_{\rm c}$} & \multicolumn{2}{c}{$K_{\rm
    v}\rightarrow K_{\rm c}$} \\

\hline

SJ  &~~~$0$&$.04$ & $-350$&$.716(1)$ &~~~~~~$1$&$.18(3)$ &~~~~~~~$4$&$.21(3)$
&~$6$&$.12(1)$ &~$6$&$.25(2)$ \\

SJ  & $0$&$.01$ & $-350$&$.739(3)$ & $1$&$.06(6)$  & $4$&$.22(6)$ &
$6$&$.09(3)$ & $6$&$.28(3)$ \\

SJ & $0$&     & $-350$&$.747(4)$ & $1$&$.02(9)$ & $4$&$.22(8)$ &
$6$&$.08(4)$ & $6$&$.29(4)$ \\

SJB & $0$&$.04$ & $-350$&$.835(3)$ & $1$&$.28(5)$ & $4$&$.28(4)$ &
$6$&$.17(3)$ & $6$&$.28(3)$ \\

SJB & $0$&$.01$ & $-350$&$.852(1)$ & $0$&$.97(3)$ & $4$&$.14(2)$ &
$6$&$.07(2)$ & $6$&$.24(2)$ \\

SJB & $0$&    & $-350$&$.857(2)$ & $0$&$.86(4)$ & $4$&$.09(4)$ &
$6$&$.04(2)$ & $6$&$.22(2)$ \\

\hline \hline
\end{tabular}
\end{center}
\end{table*}

Apart from these tests we have used the ground-state-optimized Jastrow factor
(and backflow function, where applicable) in all our excited-state
calculations.  The fixed-node SJ-DMC energy does not depend on the
Jastrow factor, except via the pseudopotential locality
approximation \cite{Mitas_1991}, and so reoptimizing the Jastrow factor
in each excited state would be pointless in any case.  The
single-particle bands at $K$ and $K^\prime$ are degenerate, and hence
we can construct multideterminant excited-state wave functions from
the degenerate orbitals.  We discuss this in
Sec.\ \ref{sec:multidets}.

\begin{table*}[!htbp]
\begin{center}
\caption{``Hartree-Fock'' VMC (HFVMC), SJ-VMC, SJB-VMC, SJ-DMC, and
  SJB-DMC ground-state (GS) total energies, energy variances,
  quasiparticle (QP) gaps, and excitonic gaps for a $3\times 3$
  supercell of monolayer hBN\@. The fact that the excitonic gap is
  higher than the quasiparticle gap is a manifestation of finite-size
  error in the uncorrected gaps, as explained in
  Sec.\ \ref{sec:finite_size}. \label{table:hfvmc_vmc_dmc_gaps}}
\begin{tabular}{lr@{}lr@{}lr@{}lr@{}lr@{}lr@{}l}
\hline \hline

& \multicolumn{2}{c}{GS energy $E^{\rm GS}$} &
\multicolumn{2}{c}{Var.\ $\sigma^2$} & \multicolumn{4}{c}{QP gap
  $\Delta_{\rm qp}$ (eV)} & \multicolumn{4}{c}{Ex.\ gap $\Delta_{\rm
    ex}$ (eV)} \\

\raisebox{1.5ex}[0pt]{Method} & \multicolumn{2}{c}{(eV/p.\ cell)} &
\multicolumn{2}{c}{(a.u.)} & \multicolumn{2}{c}{$K_{\rm v}\rightarrow
  \Gamma_{\rm c}$} & \multicolumn{2}{c}{$K_{\rm v}\rightarrow K_{\rm
    c}$} & \multicolumn{2}{c}{$K_{\rm v}\rightarrow \Gamma_{\rm c}$} &
\multicolumn{2}{c}{$K_{\rm v}\rightarrow K_{\rm c}$} \\

\hline

HFVMC & $-341$&$.961(4)$ & $21$&$.39$ &~$2$&$.63(8)$ &~$5$&$.95(8)$
&~$7$&$.13(5)$ &~$7$&$.65(5)$ \\

SJ-VMC & $-349$&$.8780(4)$ & $3$&$.18$ & $2$&$.559(9)$ & $4$&$.593(9)$
& $7$&$.118(6)$ & $6$&$.378(5)$ \\

SJB-VMC & $-350$&$.229(2)$ & $2$&$.11$ & $2$&$.55(4)$ & $4$&$.46(4)$
&$7$&$.18(2)$ & $6$&$.30(2)$ \\

SJ-DMC & $-350$&$.747(4)$ & & & $1$&$.02(9)$ &
$4$&$.22(8)$ & $6$&$.08(4)$ & $6$&$.29(4)$ \\

SJB-DMC & $-350$&$.857(2)$ & & & $0$&$.86(4)$ &
$4$&$.09(4)$ & $6$&$.04(2)$ & $6$&$.22(2)$ \\

\hline \hline
\end{tabular}
\end{center}
\end{table*}

\subsubsection{DMC time step, etc.}

The time-step error in the total energy per primitive cell is clearly
significant, as shown in Table \ref{table:dmc_energies}; however,
there is a partial cancellation of time-step errors when we take
differences of total energies to obtain gaps.  For the SJ-DMC gaps the
time-step errors are of marginal significance.  Nevertheless, since we
would like to achieve very high accuracy, we have used DMC time steps
of 0.01 a.u.\ and 0.04 a.u.\ and extrapolated our results linearly to
zero time step.  The time-step errors in our SJB-DMC gaps are
considerably larger.  All our DMC calculations used populations of at
least 1024 walkers, making population-control bias negligible.  We
used Dirac-Fock pseudopotentials \cite{Trail_2005a,Trail_2005b} to
represent the boron and nitrogen atoms, including core-polarization
corrections \cite{Shirley_1993}.

\subsubsection{Comparison of VMC and DMC gap results}

VMC is considerably cheaper than DMC, typically by a factor of at
least ten.  VMC can therefore be used to study larger systems than
DMC\@.  However, whereas fixed-node DMC total energies and band gaps
are independent of the Jastrow factor in the limit of zero time step
and large population, VMC energies are determined by the Jastrow
factor.  The use of a stochastically optimized Jastrow factor is
therefore an additional source of noise in the VMC gaps.  VMC and DMC
results obtained with different levels of trial wave function for a $3
\times 3$ supercell are presented in Table
\ref{table:hfvmc_vmc_dmc_gaps}. The trial wave function in the
``Hartree-Fock'' VMC calculations was simply a Slater determinant of
DFT-PBE orbitals, with no description of correlation.  The fractions
of correlation energy retrieved at the SJ-VMC and SJB-VMC levels are
clearly different in the ground state and excited states.  However, we
find the VMC gaps to be larger than the DMC gaps, and the SJ gaps to
be larger than the SJB ones, as expected.  We do not believe our VMC
results can be used to aid the extrapolation of our DMC gaps to the
thermodynamic limit of infinite system size.

\subsubsection{Singlet and triplet excitations\label{sec:spin_states}}

We have calculated the SJ-DMC energy difference between the singlet
and triplet excitonic states in a $3 \times 3$ supercell of hBN\@.  We
used single-determinant trial wave functions in which an electron was
promoted with and without a spin-flip to describe the triplet and
singlet states, respectively \cite{Marsusi_2011}. The orbitals were
generated in a non-spin-polarized ground-state DFT-PBE calculation.
The singlet excitonic state for a promotion from $K_{\rm v}
\rightarrow K_{\rm c}$ is 0.12(2) eV lower in energy than the triplet
state.  For a promotion from $K_{\rm v} \rightarrow \Gamma_{\rm c}$,
the energy difference between the singlet and triplet excitonic states
is statistically insignificant (smaller than the error bar of 0.02
eV)\@.  Because these results were obtained in a small ($3\times 3$)
supercell, these estimates of the singlet-triplet splitting should
only be regarded as being of qualitative accuracy.

In summary, the energy difference between the singlet and triplet
excitonic states in hBN appears to be small, especially when the
electron and hole have different wave vectors.  Apart from these
tests, all the exited-state calculations reported in this paper used
singlet excitations.

\subsubsection{Multideterminant wave functions\label{sec:multidets}}

We have considered three different ways of describing the wave
function of a singlet excitonic state: (i) simply promoting a single
electron from one state to another without changing its spin in a
single-determinant wave function; (ii) constructing a two-determinant
wave function in which spin-up and spin-down electrons are promoted in
the first and second determinants, respectively; and (iii)
constructing a multideterminant wave function consisting of a linear
combination of all the degenerate excited-state determinants (i.e.,
accounting for the degeneracy of $K$ and $K^\prime$ as well as
spin-degeneracy).  In case (iii), we have 4- and 8-determinant wave
functions for excitations from $K_{\rm v} \rightarrow \Gamma_{\rm c}$
and from $K_{\rm v} \rightarrow K_{\rm c}$, respectively.  We
optimized the determinant expansion coefficients for these
multideterminant wave functions in the presence of a fixed Jastrow
factor that was optimized in the ground state, but we did not find a
statistically significant reduction in the VMC energy.  Any reduction
in the DMC energy would be even smaller and hence we conclude that, to
the level of precision at which we are working, there is no advantage
to using such a multideterminant wave function.  This does not imply
that larger multideterminant wave functions would not significantly
reduce fixed-node errors.

\subsubsection{Finite-size effects\label{sec:finite_size}}

We have performed QMC calculations for monolayer hBN in a range of
hexagonal supercells, from $2\times 2$ to $9\times 9$ primitive cells.
Choosing the monolayer supercell to be hexagonal maximizes the
distance between nearest periodic images of particles, and is
therefore expected to minimize finite-size effects.  For bulk hBN the
choice of supercell is more complicated.  In general a supercell is
defined by an integer ``supercell matrix'' $S$ such that
\begin{equation}
  {\bf a}^{\rm S}_{i} = \sum_{k}S_{ik}{\bf a}^{\rm P}_{k},
\end{equation}
where ${\bf a}^{\rm P}_{k}$ is the $k^{\text{th}}$ primitive lattice
vector and ${\bf a}^{\rm S}_{i}$ is the $i^{\text{th}}$ supercell
lattice vector.  The supercell defined by $S$ contains $N_{\rm
  P}=|\det(S)|$ primitive cells.  For a given number of primitive
cells $N_{\rm P}$ we may therefore search over integer supercell
matrices $S$ such that $|\det(S)|=N_{\rm P}$ to find the supercell
matrix for which the nearest-image distance is maximized
\cite{LloydWilliams_2015}.  In general the optimal supercell matrix is
nondiagonal. Our bulk hBN supercells contained $N_{\rm P}=9$, $18$,
$27$, and $36$ primitive cells.  Unlike the monolayer, in bulk hBN we
are unable to choose a large set of geometrically similar supercells
that both maximize the distance between periodic images and have a
tractable number of particles.

Different choices of supercell Bloch vector
\cite{Rajagopal_1994,Rajagopal_1995} allow one to obtain different
points on the electronic band structure in a finite supercell
\cite{Mitas_1994,Williamson_1998,Towler_2000}. In the monolayer, if
one uses a $3m \times 3n$ supercell, where $m$ and $n$ are natural
numbers, with the supercell Bloch vector being ${\bf k}_{\rm s}={\bf
  0}$, then the set of orbitals in the trial wave function includes
the bands at both the $\Gamma$ and the $K$ points of the
primitive-cell Brillouin zone.  In this case one can make additions or
subtractions at $\Gamma$ or $K$ and promote electrons either from
$K_{\rm v} \rightarrow \Gamma_{\rm c}$ or from $K_{\rm v} \rightarrow
K_{\rm c}$.  In a general supercell, however, one can choose the
supercell Bloch vector ${\bf k}_{\rm s}$ so that the orbitals at
$\Gamma$ are present in the Slater wave function, or so that the
orbitals at $K$ are present, but not both at the same time.  The
quasiparticle gap from $K_{\rm v} \rightarrow \Gamma_{\rm c}$ can
always be calculated for a given supercell by determining the CBM and
VBM using two different values of ${\bf k}_{\rm s}$. Similar comments
apply in the bulk case.  With the optimal (nondiagonal) supercell
matrices for $N_{\rm P}=\det{(S)}=18$ or $36$ primitive cells we are
unable to include $\Gamma$ and $K$ simultaneously in the grid of ${\bf
  k}$ vectors. This prevents calculation of the $\Gamma_{\rm v}
\rightarrow K_{\rm c}$ excitonic gap in these supercells. We have
instead calculated the bulk $\Gamma_{\rm v} \rightarrow K_{\rm c}$
excitonic gaps in supercells defined by diagonal supercell matrices
$S(N_{\rm P}=18)=\text{diag}(3,3,2)$ and $S(N_{\rm
  P}=36)=\text{diag}(3,3,4)$.

Now let us consider ``long-range'' finite-size errors in the energy
gaps in periodic supercells.  Adding a single electron to or removing
a single electron from a periodic simulation supercell results in the
creation of an unwanted lattice of quasiparticles at the set of
supercell lattice points \cite{Hunt_2018}. The leading-order systematic
finite-size error in the quasiparticle bands is therefore $v_{\rm
  M}/2$, where $v_{\rm M}$ is the screened Madelung
constant \cite{Madelung_1918} of the supercell. The situation is
qualitatively similar to that encountered in \textit{ab initio}
simulations of charged defects \cite{Hine_2009}. Following the notation
of Sec.\ \ref{sec:eval_gaps}, a finite-size-corrected expression for
an unoccupied energy band is ${\cal E}_i'({\bf k})=[E^+({\bf
    k},i)-v_{\rm M}/2]-E^{\rm GS}$.  In a similar fashion, when one
creates a lattice of holes by removing an electron from a periodic
supercell, a finite-size-corrected expression for an occupied energy
band is ${\cal E}_i'({\bf k})=E^{\rm GS}-[E^-({\bf k},i)-v_{\rm
    M}/2]$.  The finite-size-corrected quasiparticle band gap is
therefore
\begin{equation} \Delta_{\rm qp}'={\cal E}_{\rm CBM}'-{\cal E}_{\rm
VBM}'=E^+_{\rm CBM}+E^-_{\rm VBM}-2E^{\rm GS}-v_{\rm
    M}. \label{eq:qp_gap_corr} \end{equation}

In an hBN monolayer, in-plane screening modifies the form of the
Coulomb interaction between charges.  The screened interaction is
approximately of Keldysh form \cite{Keldysh_1979}. Including the
relative permittivity $\epsilon$ of the surrounding medium, the
Keldysh interaction in reciprocal space is $v(k)=2\pi/[\epsilon
  k(1+r_*k)]$, where $r_*$ is the ratio of the in-plane susceptibility
of the layer to the permittivity of the surrounding medium. For the
monolayer the leading-order finite-size error in the
quasiparticle gap is the Madelung constant $v_{\rm M}$ of the
supercell \textit{evaluated using the Keldysh
  interaction} \cite{Hunt_2018}. In small supercells, the Keldysh
interaction between nearby periodic images varies logarithmically with $r$
and the Madelung constant is almost independent of system size;
however, once the supercell size significantly exceeds
$r_*$, the Keldysh interaction reduces to the Coulomb $1/r$ form and the
Madelung constant falls off as the reciprocal of the linear size of
the supercell.

The parameter $r_*$ may be estimated using Eq.\ (S.7) in the
Supplemental Material of Ref.\ \onlinecite{Ganchev_2015}.  For
free-standing monolayer hBN, $r_* \approx c(\epsilon_\|-1)/4$, where
$c$ is the out-of-plane lattice parameter of bulk hBN, $\epsilon_\|$
is the in-plane component of the high-frequency permittivity tensor,
and we have included an extra factor of $1/2$ due to the fact that
there are two layers per bulk hBN primitive cell. The lattice
parameter is measured to be $c=6.6612$ {\AA} \cite{Lynch_1966}, while
DFT-PBE calculations predict that $\epsilon_\| \approx 4.69$, giving
$r_* \approx 6.14$ {\AA}.  Unfortunately the supercell sizes used in
this work are comparable in size to $r_*$.  For bulk hBN the $r_*$
value is smaller by a factor $\sqrt{\epsilon_\| \epsilon_z}$, where
$\epsilon_z$ is the out-of-plane component of the permittivity
tensor \cite{Ganchev_2015}. Using the DFT-PBE high-frequency
out-of-plane permittivity $\epsilon_z=2.65$ gives $r_*^{\rm
  bulk}=1.74$ {\AA} for bulk hBN\@.  Our bulk hBN supercells are
sufficiently large that the interaction between periodic images can be
assumed to be of Coulomb $1/r$ form; nevertheless, the strong
anisotropy of the dielectric screening must be taken into account in
the evaluation of the screened Madelung constant \cite{Murphy_2013}
$v_{\rm M}$.

The remaining systematic finite-size effects in the quasiparticle gap
are primarily due to charge-quadrupole image interactions and fall off
as $a_{\rm s}^{-2}$ when $a_{\rm s} \leq r_*$ and as $a_{\rm s}^{-3}$
when $a_{\rm s} \gg r_*$, where $a_{\rm s}$ is the in-plane linear
size of the supercell \cite{Hunt_2018}. There are also oscillatory,
quasirandom errors with a slowly decaying envelope as a function of
system size due to long-range oscillations in the pair-correlation
function being forced to be commensurate with the supercell (see
Figs.\ \ref{fig:gap_v_NP} and \ref{fig:bulk_fs}).  We remove the
remaining finite-size errors by extrapolating the Madelung-corrected
quasiparticle gaps in supercells of 9 or more primitive cells to
infinite system size, assuming the finite-size error decays as $a_{\rm
  s}^{-2}$ in monolayer hBN and as $a_{\rm s}^{-3}$ in bulk hBN (i.e.,
as $N_{\rm P}^{-1}$ in both cases).  Since the quasirandom finite-size
errors dominate the QMC statistical error bars, we do not weight our
data by the QMC error bars \cite{Hunt_2018}.

Note that the uncorrected quasiparticle gap calculated in a finite
supercell may be smaller than the excitonic gap in that supercell due
to a negative Madelung constant, as can be seen in Tables
\ref{table:dmc_energies} and \ref{table:hfvmc_vmc_dmc_gaps}.  This is
simply an artifact of the use of periodic boundary conditions and the
Ewald interaction, and the effect disappears in the thermodynamic
limit of infinite system size, where the excitonic gap must always be
less than or equal to the quasiparticle gap due to the attractive
interaction between electrons and holes.

Finite-size effects in DMC excitonic gaps may arise from the
confinement of a neutral exciton in a periodic simulation
supercell. Once the supercell size significantly exceeds the size of
the exciton, the exciton wave function is exponentially localized
within the supercell; however, power-law finite-size effects in the
exciton binding energy remain due to the difference between the
screened Coulomb interaction in a finite, periodic supercell and in an
infinite system.  The length scale of an exciton under the Keldysh
interaction is $r_0=\sqrt{r_*/(2\mu)}$, where $\mu$ is the
electron-hole reduced mass \cite{Ganchev_2015,Mostaani_2017}. Using
the DFT-HSE06 effective masses in Table \ref{table:dft_eff_masses}
together with the $r_*$ value estimated above, we find the sizes of
both the $K_{\rm v}\rightarrow\Gamma_{\rm c}$ and $K_{\rm
  v}\rightarrow K_{\rm c}$ excitons in monolayer hBN to be $r_0
\approx 3$ {\AA}, which is only slightly larger than the lattice
constant. Hence all our simulation supercells are large enough to
contain the excitons, and so the remaining finite-size effects are due
to instantaneous dipole-dipole interactions between identical images,
evaluated with the Keldysh interaction in the case of the
monolayer. On the other hand, the fact that the exciton radius is
comparable with the lattice constant implies that we are at the limit
of the validity of the effective-mass model of excitons and it may not
fully account for finite-size effects in gaps; nevertheless, this model
provides us with best available framework for understanding systematic
finite-size effects.  In supercells with $a_{\rm s} \leq r_*$, the
leading-order systematic finite-size effects in the excitonic gap go
as $a_{\rm s}^{-2}$; for supercells with $a_{\rm s} \gg r_*$, the
finite-size effects go as $a_{\rm s}^{-3}$.  We therefore extrapolate
our uncorrected excitonic gaps in the same way that we extrapolate our
Madelung-corrected quasiparticle gaps to infinite system size, i.e.,
assuming the errors go as $N_{\rm P}^{-1}$ for both bulk and
monolayer.  Again we do not weight our data by the QMC error bars,
since the quasirandom finite-size effects dominate the QMC error bars.

Our DMC gaps against system size are presented and discussed in
Sec.\ \ref{sec:finite_size_results}.

\subsection{Vibrational contribution} \label{subsec:vib_method}

We calculated the vibrational contribution to the quasiparticle band
gap arising from the electron-phonon interaction at temperature $T$
within the Born-Oppenheimer approximation as:
\begin{equation}
\Delta_{\rm
  qp}(T)=\frac{1}{\mathcal{Z}}\sum_{\mathbf{s}}\langle\Phi_{\mathbf{s}}(\mathbf{u})|\Delta_{\rm
  qp}(\mathbf{u})|\Phi_{\mathbf{s}}(\mathbf{u})\rangle
e^{-E_{\mathbf{s}}/(k_{\mathrm{B}}T)} \label{eq:gap_temperature}
\end{equation}
where the harmonic vibrational wave function
$|\Phi_{\mathbf{s}}(\mathbf{u})\rangle$ in state $\mathbf{s}$ has
energy $E_{\mathbf{s}}$, $\mathbf{u}=\{u_{\nu\mathbf{q}}\}$ is a
collective coordinate for all the nuclei written in terms of normal
modes of vibration $(\nu,\mathbf{q})$,
$\mathcal{Z}=\sum_{\mathbf{s}}e^{-E_{\mathbf{s}}/(k_{\mathrm{B}}T)}$ is
the partition function, and $k_{\mathrm{B}}$ is Boltzmann's constant.

We evaluated Eq.\ (\ref{eq:gap_temperature}) using two complementary
methods recently reviewed in Ref.\ \onlinecite{Monserrat_2018}. The
first relies on a stochastic Monte Carlo sampling of the vibrational
density over $M$ points:
\begin{equation}
\Delta_{\rm qp}^{\rm MC}(T)=\frac{1}{M}\sum_{i=1}^M\Delta_{\rm
  qp}(\mathbf{u}_i), \label{eq:temperature_mc}
\end{equation}
where configurations $\mathbf{u}_i$ are distributed according to the
nuclear density. This approach enables the inclusion of the
electron-phonon interaction at all orders at the expense of using
large diagonal supercell matrices, and in practice we use thermal
lines to accelerate the sampling \cite{Monserrat_2016b}. The second
approach relies on a second order expansion of the dependence of
$\Delta_{\rm qp}(\mathbf{u})$ on the mode amplitudes $\mathbf{u}$,
which leads to a particularly simple \textit{quadratic} approximation:
\begin{equation}
\Delta_{\rm qp}^{\rm quad}(T)=\Delta_{\rm
  qp}+\frac{1}{N_{\mathbf{q}}}\sum_{\mathbf{q},\nu}\frac{1}{\omega_{\mathbf{q}\nu}}\frac{\partial^2\Delta_{\rm
    qp}}{\partial
  u^2_{\mathbf{q}\nu}}\left[\frac{1}{2}+n_{\mathrm{B}}(\omega_{\mathbf{q}\nu},T)\right], \label{eq:temperature_quadratic}
\end{equation}
where $n_{\mathrm{B}}(\omega_{\mathbf{q}\nu},T)$ is a Bose-Einstein
factor. This expression can be efficiently evaluated using nondiagonal
supercell matrices \cite{LloydWilliams_2015} at the expense of
neglecting higher-order terms in the electron-phonon
interaction. Overall, Eq.\ (\ref{eq:temperature_quadratic}) enables
the convergence of the calculations with respect to supercell size (or
equivalently $\mathbf{q}$-point grid density), whereas
Eq.\ (\ref{eq:temperature_mc}) enables the inclusion of higher-order
terms, which have been found to provide important contributions in a
range of materials \cite{Monserrat_2015,Saidi_2016}.

All our vibrational calculations were performed using the PBE
functional, an energy cutoff of $700$ eV, and a $\mathbf{k}$-point
spacing of $2\pi\times0.025$\AA$^{-1}$ to sample the electronic
Brillouin zone. The results show slow convergence with respect to the
$\mathbf{q}$-point grid size: the vibrational correction to the
quasiparticle gap at $300$ K using the expression in
Eq.\ (\ref{eq:temperature_quadratic}) converges to values better than
$0.05$ eV using a grid size of $32\times32$ $\mathbf{q}$-points for
the monolayer, and using a grid size of $16\times16\times16$ for the
bulk. We also tested the inclusion of van der Waals dispersion
corrections in the bulk calculations using the Tkatchenko-Scheffler
scheme \cite{Tkatchenko_2009} but found differences smaller than $0.01$
eV compared to the calculations without dispersion corrections. Using
Eq.\ (\ref{eq:temperature_mc}) instead of
Eq.\ (\ref{eq:temperature_quadratic}) leads to a significant
enhancement to the vibrational correction to the quasiparticle
gap. However, calculations using Eq.\ (\ref{eq:temperature_mc}) are
restricted to smaller $\mathbf{q}$-point grid sizes, and therefore our
final results were estimated by using the $\mathbf{q}$-point converged
results obtained with Eq.\ (\ref{eq:temperature_quadratic}) and adding
a correction equal to $\Delta_{\rm qp}^{\rm MC}(T)-\Delta_{\rm
  qp}^{\rm quad}(T)$ evaluated at the largest $\mathbf{q}$-point grid
size feasible within the Monte Carlo method, which is $8\times8$ for
the monolayer and $4\times4\times4$ for the bulk.

\section{Results\label{sec:results}}

\subsection{Lattice parameter and dynamical
  stability\label{sec:geom_and_phonons}}

The lattice parameters obtained in DFT-LDA, DFT-PBE, and DFT-HSE06
calculations are $a=2.491$, $2.512$, and 2.45 {\AA}, respectively,
which may be compared with the bulk lattice parameter $a=2.5040$
{\AA} \cite{Lynch_1966} and the lattice parameter 2.5 {\AA} measured in
a thin film of hBN \cite{Shi_2010}. Our DFT-PBE lattice parameter is
in good agreement with a previously published result, $a=2.51$
{\AA} \cite{Ribeiro_2011}. We have used the DFT-PBE lattice parameter
$a=2.512$ {\AA} in all our QMC calculations.  The partial charge of
each boron atom is 0.83 according to Mulliken population analysis of the
DFT orbitals \cite{Mulliken_1955} and 0.21 a.u.\ according to Hirshfeld
analysis of the charge density \cite{Hirshfeld_1977}. The partial
charges predicted by the LDA and PBE functionals agree.

\begin{figure}[!htbp]
\begin{center}
\includegraphics[clip,width=0.45\textwidth]{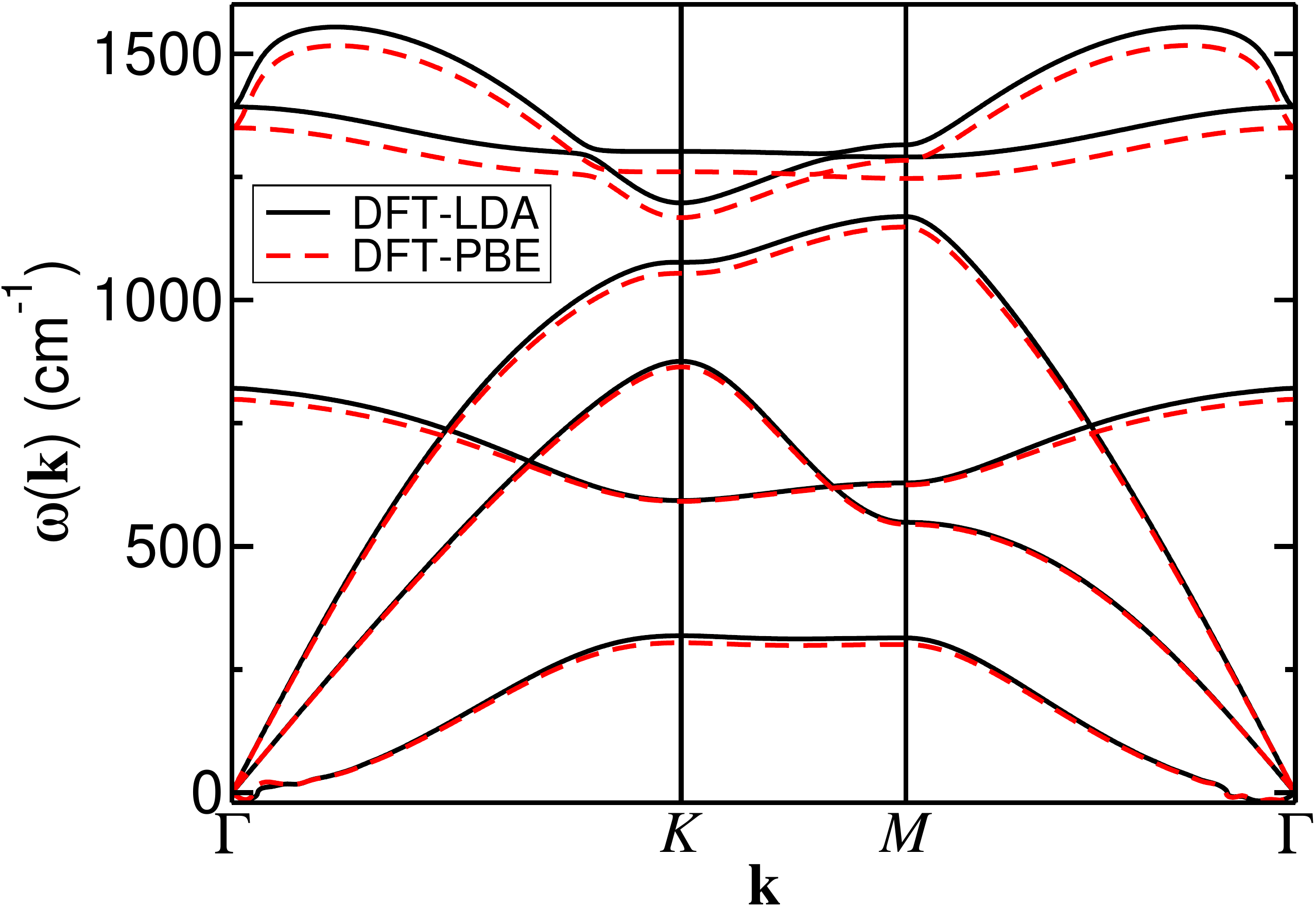}
\caption{(Color online) DFT-LDA and DFT-PBE phonon dispersion curves
  for monolayer hBN\@.
\label{fig:bn_dft_phonons}}
\end{center}
\end{figure}

The DFT-LDA and DFT-PBE phonon dispersion curves of hBN are shown in
Fig.\ \ref{fig:bn_dft_phonons}.  The calculations appear to predict a
small region of dynamical instability in the flexural acoustic branch
about the $\Gamma$ point.  Such regions of instability around $\Gamma$
are a common feature in first-principles lattice-dynamics calculations
for 2D materials, including graphene, molybdenum disulfide, and indium
and gallium chalcogenides \cite{Zolyomi_2014}. We observe that (i) the
region of instability occurs in both finite-displacement (supercell)
calculations and in density functional perturbation theory
calculations; (ii) the region of instability depends sensitively on
every simulation parameter (basis set, ${\bf k}$-point sampling,
supercell size in finite-displacement calculations,
exchange-correlation functional, pseudopotential, and artificial
periodicity); (iii) the size of the instability is the same as the
amount by which the acoustic branches miss zero at $\Gamma$ if
Newton's third law is not imposed on the force constants; and (iv) the
region of instability remains even if the layer is put under tension
by increasing the lattice parameter slightly.  To minimize the effects
of longitudinal/transverse optic-mode splitting in our
three-dimensionally periodic calculations, we choose the $z$-component
of the wave vector to be $\pi/L$, where $L$ is the artificial
periodicity \cite{Wirtz_2003}. Our results are in good agreement with
the phonon dispersion curves obtained by Wirtz \textit{et
  al.}\ \cite{Wirtz_2003}. An analysis of the Raman activity of phonon
modes is given in Ref.\ \onlinecite{Wirtz_2005}.

\subsection{DFT electronic band structure and effective
  masses\label{sec:dft_bands_masses}}

\begin{figure}[!htbp]
\begin{center}
\includegraphics[clip,width=0.45\textwidth]{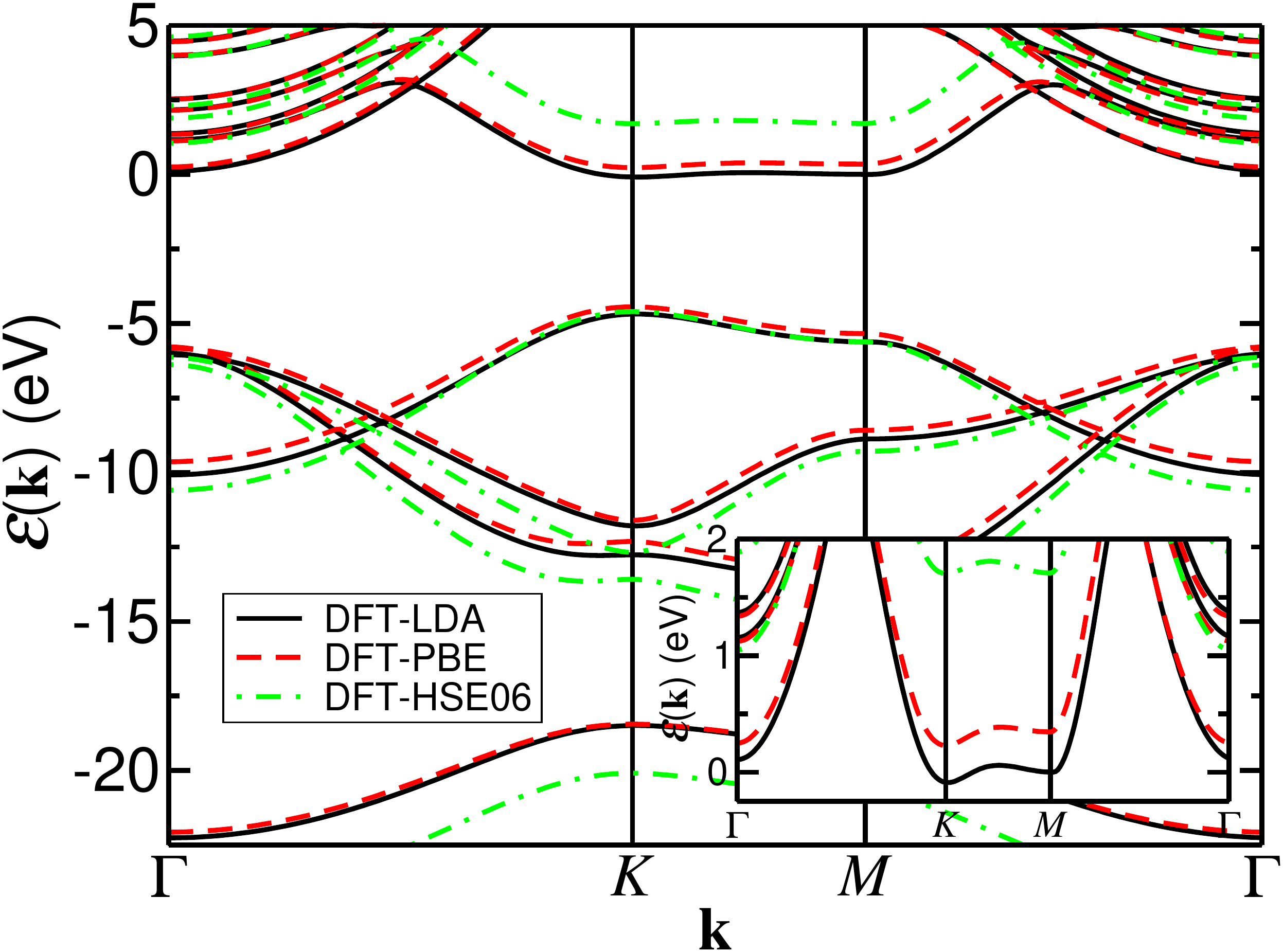}
\caption{(Color online) DFT-LDA, DFT-PBE, and DFT-HSE06 electronic
  band-structure plots for monolayer hBN\@. The zero of energy is set to the
  Fermi energy. The inset shows the energy range around the CBM in
  greater detail.
\label{fig:bn_dft_bs}}
\end{center}
\end{figure}

\begin{figure}[!htbp]
\begin{center}
\includegraphics[clip,width=0.45\textwidth]{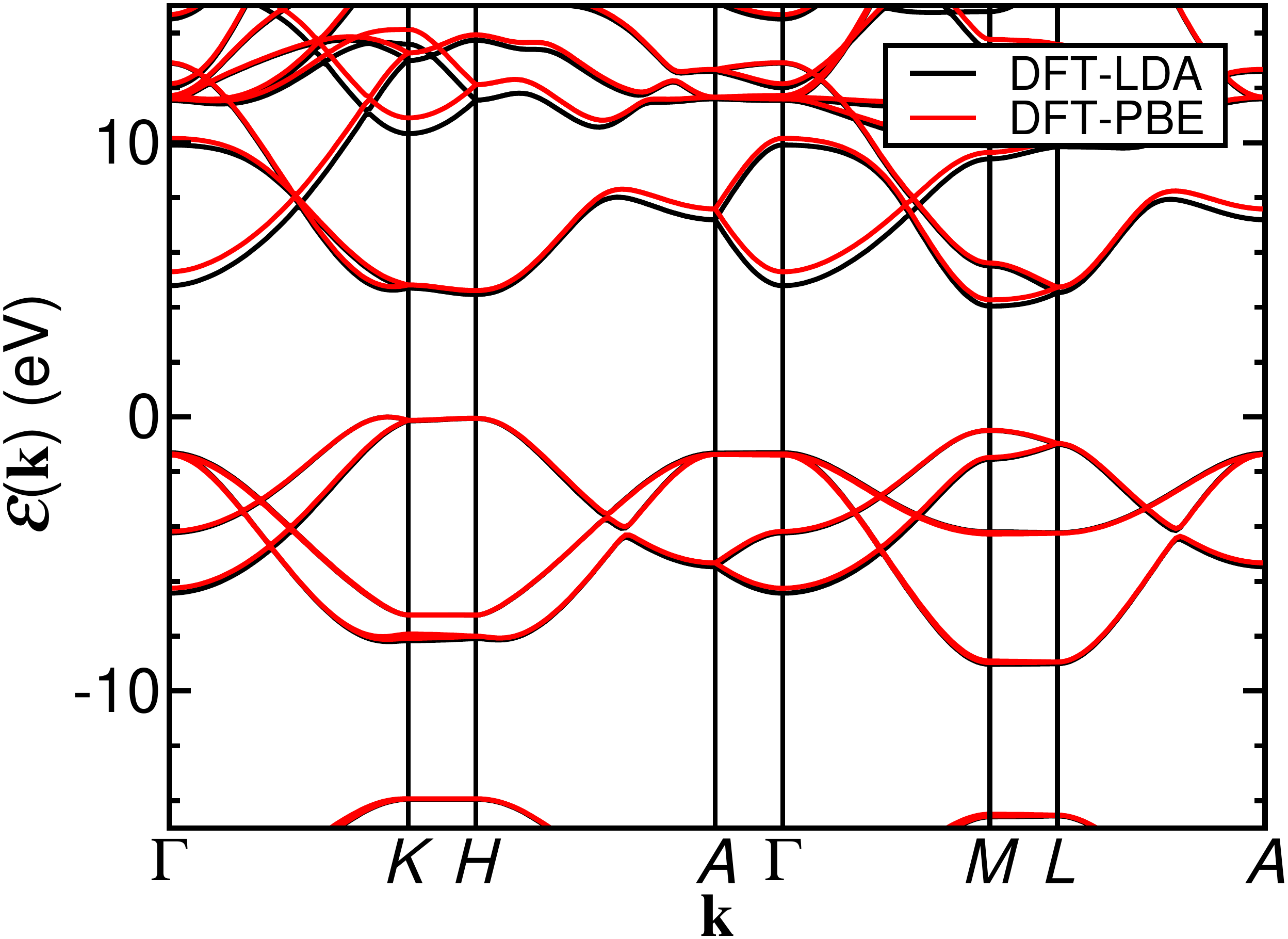}
\caption{(Color online) DFT-LDA and DFT-PBE electronic band-structure
  plots for bulk hBN\@. The zero of energy is set to the Fermi energy.
\label{fig:bulk_bn_dft_bs}}
\end{center}
\end{figure}

The DFT-LDA, DFT-PBE, and DFT-HSE06 band structures of monolayer and
bulk hBN are shown in Figs.\ \ref{fig:bn_dft_bs} and
\ref{fig:bulk_bn_dft_bs}, respectively.  In the case of the monolayer,
we fitted
\begin{eqnarray} {\cal E}_{\rm c,v}({\bf q}) & = & {\cal
    E}_{K_{\rm c,v}} \pm
  \frac{q^2}{2m_{K_{\rm c,v}}^\ast}+A_{\rm c,v}q^4+B_{\rm c,v}q^6 \nonumber \\ & &
       {}+C_{\rm c,v}q^6\cos(6\theta) +D_{\rm c,v}q^3\cos(3\theta) \nonumber
       \\ & &
          {}+E_{\rm c,v}q^5\cos(3\theta), \label{eq:band_K} \end{eqnarray}
where ${\cal E}_{K_{\rm c,v}}$, $m_{K_{\rm c,v}}^\ast$, $A_{\rm c,v}$, $B_{\rm c,v}$,
$C_{\rm c,v}$, $D_{\rm c,v}$, and $E_{\rm c,v}$ are fitting parameters, to the
conduction and valence bands within a circle of radius 6\% of the
$\Gamma$--$M$ distance around the $K$ point.  ${\bf q}$ is the
wavevector relative to the $K$ point, and $\theta$ is the polar angle
of ${\bf q}$.  The second term is positive for the conduction band and
negative for the valence band, so that $m_{K_{\rm c}}^\ast$ and
$m_{K_{\rm v}}^\ast$ are the electron and hole effective masses. The
root-mean-square (RMS) residual over the fitting area is less than 0.2
meV in each case.  We fitted
\begin{equation} {\cal E}_{\rm c}({\bf k}) = {\cal
    E}_{\Gamma_{\rm c}} + \frac{k^2}{2m_{\Gamma_{\rm
        c}}^\ast}+A^\prime k^4+B^\prime k^6+C^\prime
  k^6\cos(6\theta), \label{eq:band_Gamma} \end{equation} where
$k=|{\bf k}|$, $\theta$ is the polar angle of ${\bf k}$, and ${\cal
  E}_{\Gamma_{\rm c}}$, $m_{\Gamma_{\rm c}}^\ast$, $A^\prime$,
$B^\prime$, and $C^\prime$ are fitting parameters, to the conduction
band within a circle of radius 40\% of the $\Gamma$--$M$ distance
about $\Gamma$. The RMS residual over this area is less than 0.2 meV
in each case.  It is clearly much easier to represent the band over a
large area around $\Gamma$ than around $K$.  The fitted effective
masses in Eqs.\ (\ref{eq:band_K}) and (\ref{eq:band_Gamma}) are
reported in Table \ref{table:dft_eff_masses}.  It was verified that
the effective masses are unchanged to the reported precision when the
radius of the circle used for the fit is reduced.

\begin{table}[!htbp]
\begin{center}
\caption{Effective masses $m^\ast$ for the $\Gamma_{\rm c}$, $K_{\rm
    c}$, and $K_{\rm v}$ bands from DFT-LDA, DFT-PBE, and DFT-HSE06
  calculations.\label{table:dft_eff_masses}}
\begin{tabular}{lccc}
\hline \hline

& \multicolumn{3}{c}{$m^\ast$ (a.u.)} \\

\raisebox{1.5ex}[0pt]{Functional} & $\Gamma_{\rm c}$ & $K_{\rm c}$ &
$K_{\rm v}$ \\

\hline

LDA    & $0.96$ & $0.89$ & $0.61$ \\

PBE    & $0.95$ & $0.90$ & $0.63$ \\

HSE06  & $0.98$ & $1.07$ & $0.63$ \\

\hline \hline
\end{tabular}
\end{center}
\end{table}

\begin{figure}[!htbp]
\begin{center}
\includegraphics[clip,width=0.45\textwidth]{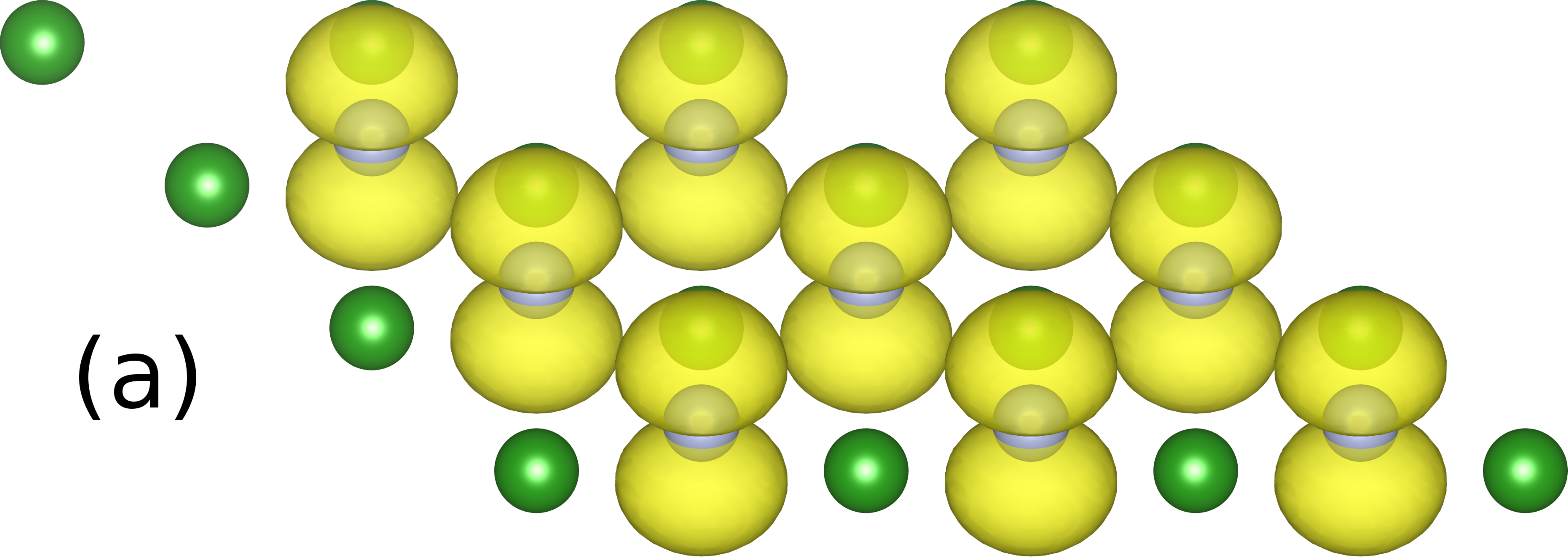} \\[1em]
\includegraphics[clip,width=0.45\textwidth]{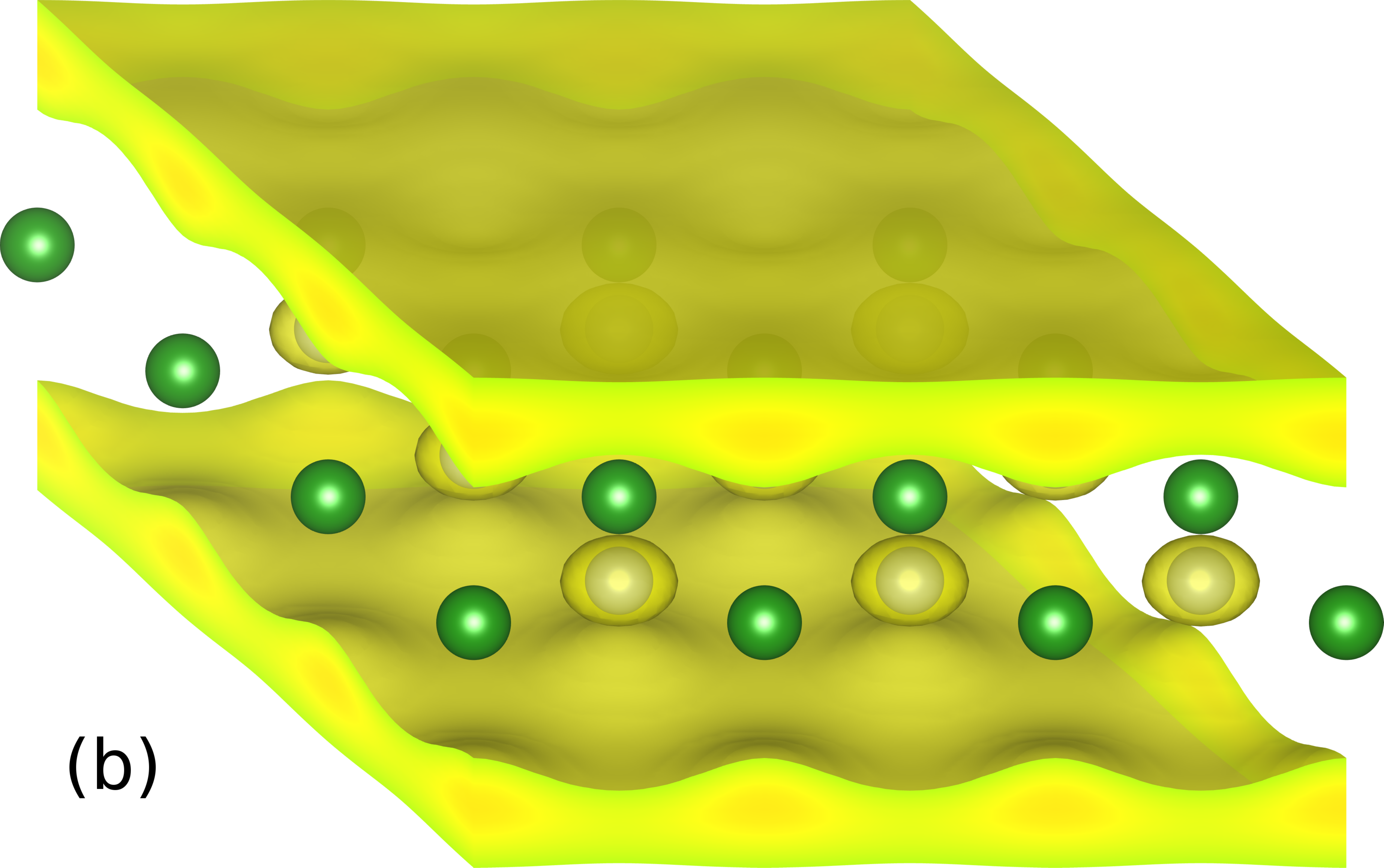} \\[1em]
\includegraphics[clip,width=0.45\textwidth]{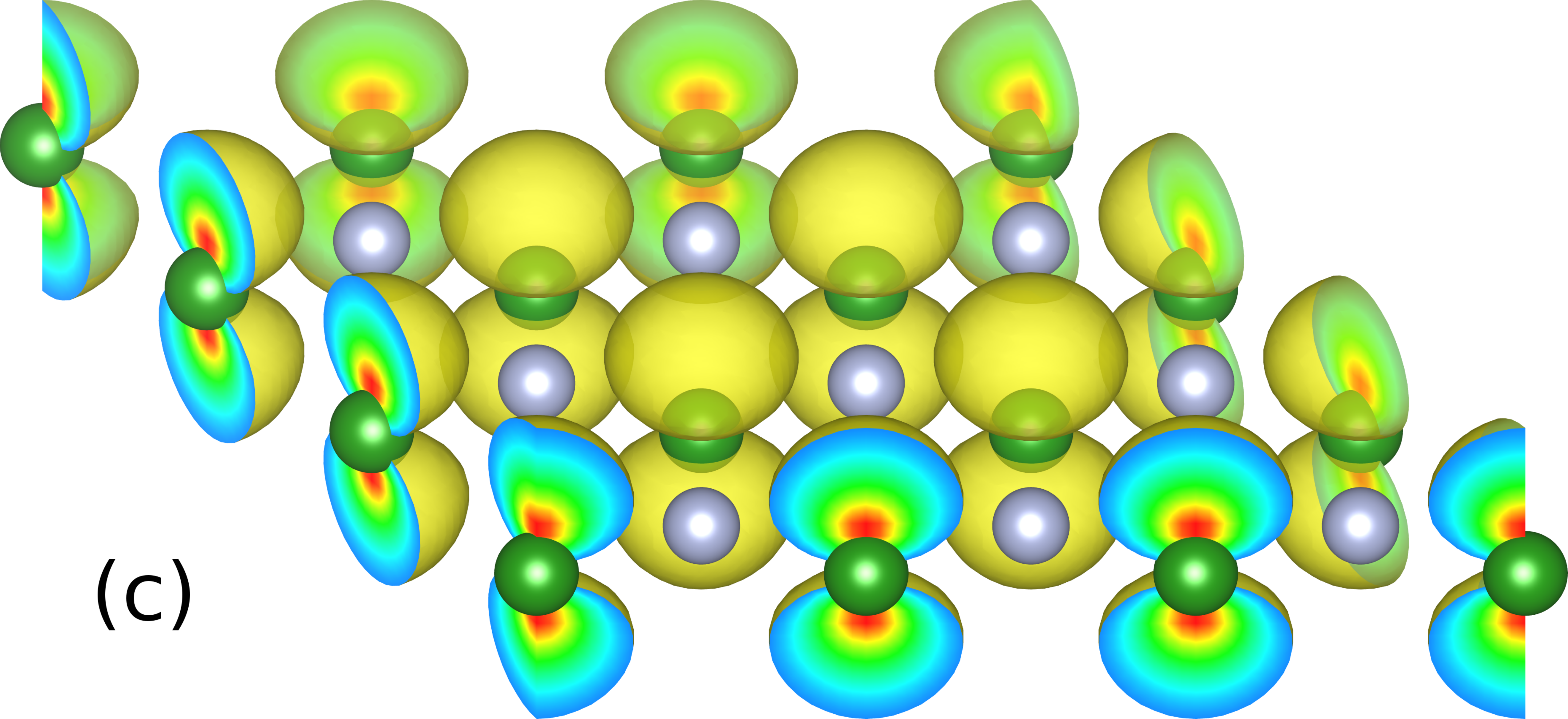}
\caption{(Color online) DFT-HSE06 charge densities of (a) the
  valence-band maximum at $K_{\rm v}$, (b) the conduction-band minimum
  at $\Gamma_{\rm c}$, and (c) the conduction-band minimum at $K_{\rm
    c}$ for monolayer hBN\@.  The green spheres show the boron atoms,
  while the white spheres are nitrogen atoms.  The charge densities
  were obtained using an artificial periodicity of 21.2 {\AA} in the
  out-of-plane direction, a $15 \times 15$ Monkhorst-Pack ${\bf
    k}$-point grid, DFT norm-conserving pseudopotentials, and a
  plane-wave cutoff energy of 680 eV\@.
\label{fig:hse_cden}}
\end{center}
\end{figure}

The DFT charge density of the conduction-band minimum at $\Gamma_{\rm
  c}$ consists of two delocalized, free-electron-like regions on
either side of the hBN layer, whereas the charge density for the
conduction-band minimum at $K_{\rm c}$ is localized on the boron
atoms: see Fig.\ \ref{fig:hse_cden}.  This is consistent with the
observation that the conduction band at $\Gamma$ is nearly parabolic
with an effective mass close to the bare electron mass. The orbital
charge densities are qualitatively similar in the monolayer and in the
bulk.

\subsection{Energy-gap results\label{sec:dmc_gap_results}}

\subsubsection{Finite-size effects in the DMC band
  gap\label{sec:finite_size_results}}

\begin{figure}[!htbp]
\begin{center}
\includegraphics[clip,width=0.45\textwidth]{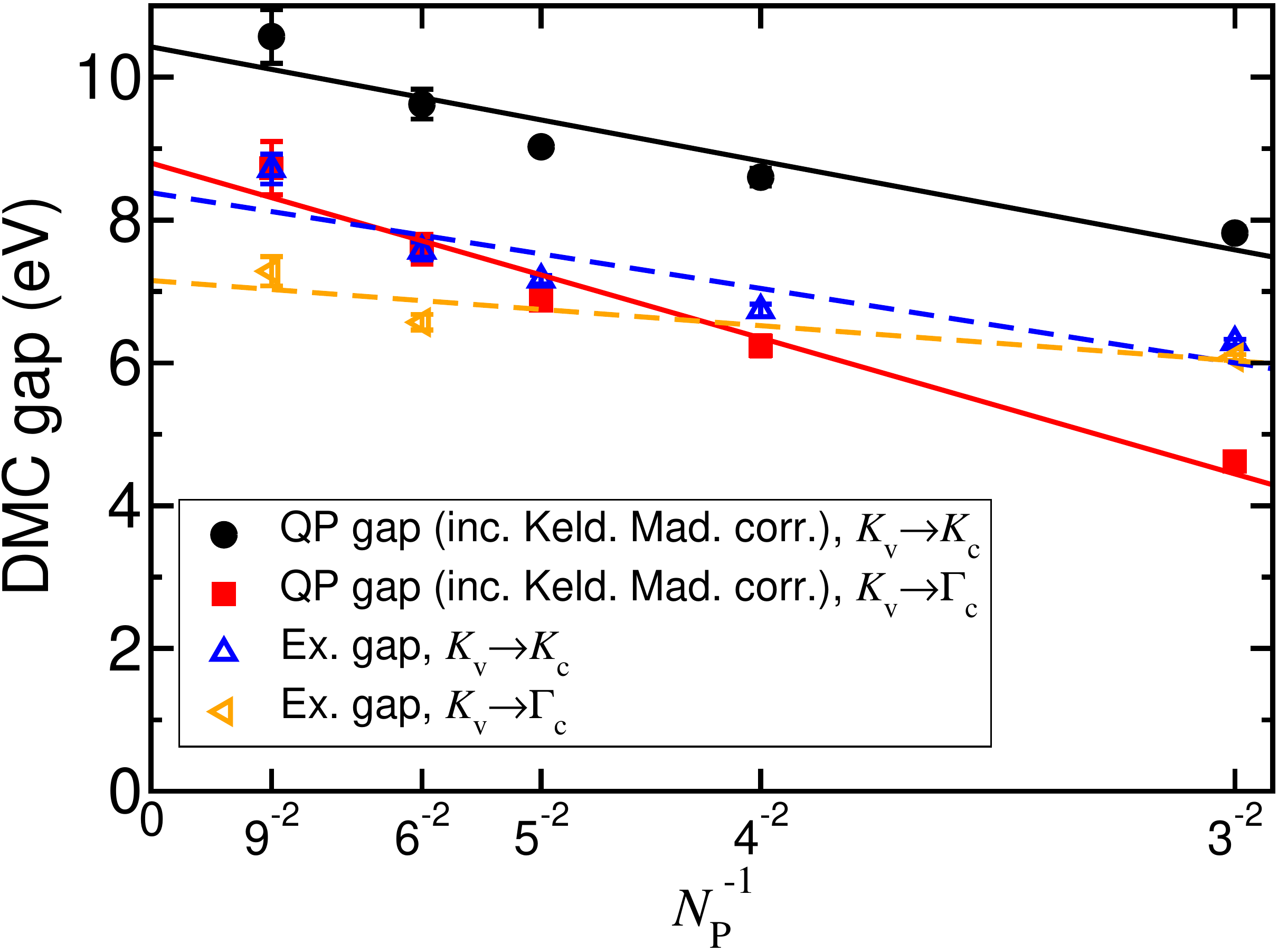}
\caption{(Color online) SJ-DMC quasiparticle (QP) and excitonic gaps of
  monolayer hBN against $N_{\rm P}^{-1}$, where $N_{\rm P}$ is the
  number of primitive cells in the supercell. The quasiparticle gaps
  include the Madelung correction given in
  Eq.\ (\ref{eq:qp_gap_corr}).
\label{fig:gap_v_NP}}
\end{center}
\end{figure}

The SJ-DMC quasiparticle and excitonic band gaps (both $K_{\rm v}
\rightarrow K_{\rm c}$ and $K_{\rm v} \rightarrow \Gamma_{\rm c}$) are
plotted against system size in Fig.\ \ref{fig:gap_v_NP}.  The
quasiparticle gaps include the correction shown in
Eq.\ (\ref{eq:qp_gap_corr}).  Systematic finite-size effects in the
$K_{\rm v} \rightarrow \Gamma_{\rm c}$ and $K_{\rm v} \rightarrow
K_{\rm c}$ excitonic gaps are very much smaller than systematic
finite-size errors in the uncorrected quasiparticle gaps.  On the
other hand the quasirandom finite-size noise in both types of gap has
an amplitude of about 0.5 eV over the range of supercells studied.

\subsubsection{Nature and size of the gap in the thermodynamic limit}

Our results for the electronic band gaps are given in Table
\ref{table:gap_results}. The error bars on our QMC gaps are determined
by the quasirandom finite-size noise discussed in
Secs.\ \ref{sec:finite_size} and \ref{sec:finite_size_results}.  The
SJ-DMC quasiparticle gap of monolayer hBN is indirect ($K_{\rm v}
\rightarrow \Gamma_{\rm c}$) and is of magnitude 8.8(3) eV, which is
considerably enhanced with respect to the gap in the bulk, and is also
significantly higher than the gap predicted by our $GW_0$ calculations
($7.72$ eV\@).

\begin{table*}[!htbp]
\begin{center}
\caption{Static-nucleus quasiparticle and excitonic gaps for monolayer
  hBN, calculated by different methods.  Our DFT calculations indicate
  that vibrational effects lead to a renormalization of the
  static-nucleus gaps by $-0.73$ eV at 300
  K\@.  \label{table:gap_results}}
\begin{tabular}{lr@{}lr@{}lr@{}lr@{}lr@{}lr@{}lr@{}l}
\hline \hline

 & \multicolumn{6}{c}{Quasiparticle gap $\Delta_{\rm qp}$ (eV)} &
\multicolumn{4}{c}{Ex.\ gap $\Delta_{\rm ex}$ (eV)} &
\multicolumn{4}{c}{Ex.\ bind.\ $\Delta_{\rm qp}-\Delta_{\rm ex}$
  (eV)} \\

\raisebox{1.5ex}[0pt]{Method} & \multicolumn{2}{c}{$K_{\rm
    v}\rightarrow \Gamma_{\rm c}$} & \multicolumn{2}{c}{$K_{\rm
    v}\rightarrow K_{\rm c}$} & \multicolumn{2}{c}{$K_{\rm
    v}\rightarrow M_{\rm c}$} & \multicolumn{2}{c}{$K_{\rm
    v}\rightarrow \Gamma_{\rm c}$} & \multicolumn{2}{c}{$K_{\rm v}
  \rightarrow K_{\rm c}$} & \multicolumn{2}{c}{$K_{\rm v}\rightarrow
  \Gamma_{\rm c}$} & \multicolumn{2}{c}{$K_{\rm v} \rightarrow K_{\rm
    c}$} \\

\hline

DFT-LDA   &~~$4$&$.79$ &~~~~~$4$&$.60$ &~~~~$4$&$.68$ & & & & & & & & \\

DFT-PBE   & $4$&$.69$ & $4$&$.67$ & $4$&$.79$ & & & & & & & & \\

DFT-HSE06 & $5$&$.65$ & $6$&$.31$ & $6$&$.31$ & & & & & & & & \\

$G_0W_0$(-BSE) & $7$&$.43$ & $7$&$.90$ & $8$&$.00$ & & & $5$&$.81$ & &
& $2$&$.09$ \\

$GW_0$(-BSE) \cite{Wirtz_2006} & & & $8$&$.2$ & & & & & ~~~~$6$&$.1$ &
& & ~~~$2$&$.1$ \\

$GW_0$(-BSE) & $7$&$.72$ & $8$&$.18$ & $8$&$.28$ & & & $6$&$.10$ & & &
$2$&$.08$ \\

SJ-DMC    & $8$&$.8(3)$  & $10$&$.4(3)$ & & &
$6$&$.9(3)$ & ~~~$8$&$.6(2)$ & ~~~$1$&$.9(4)$ & ~~~~~~$1$&$.8(4)$ \\

\hline \hline
\end{tabular}
\end{center}
\end{table*}

\begin{figure}[!htbp]
\begin{center}
\includegraphics[clip,width=0.45\textwidth]{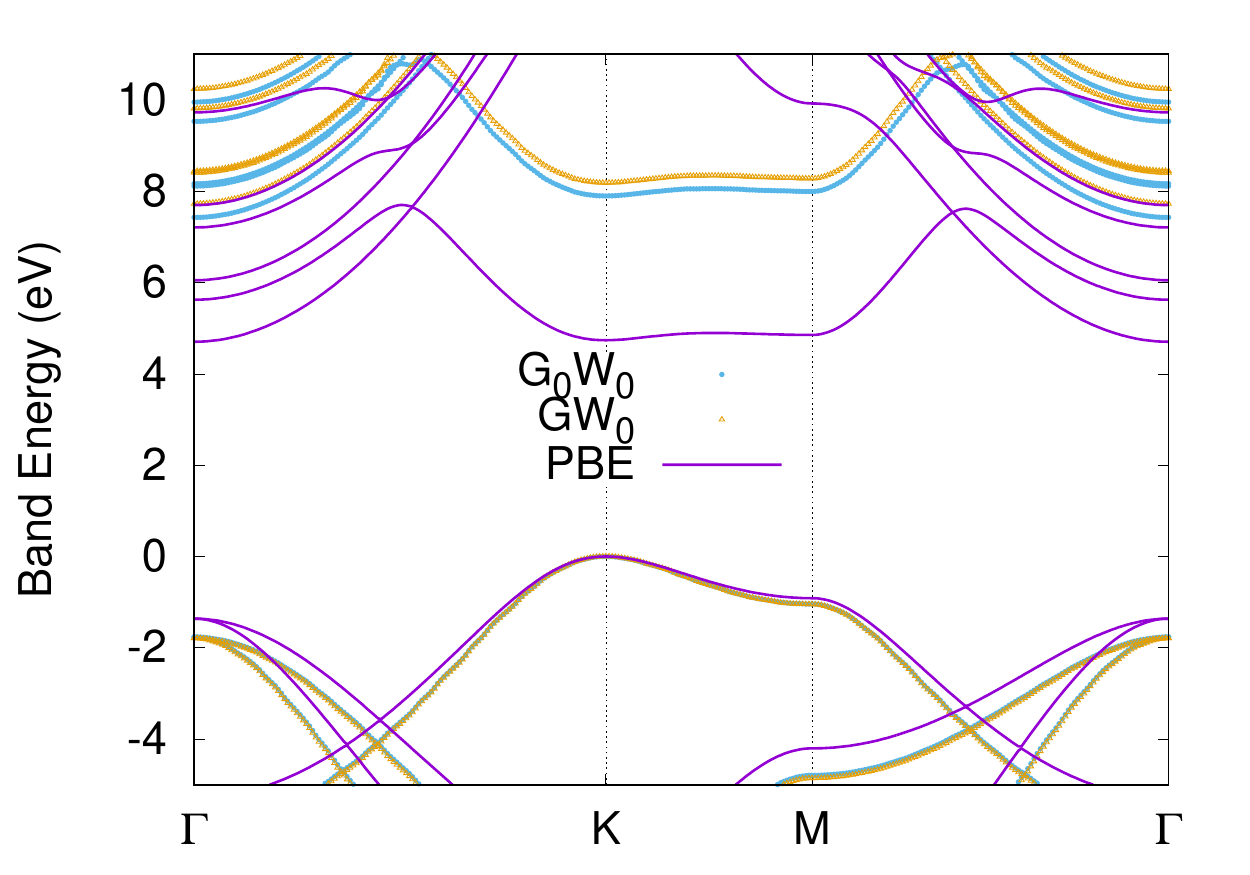} \\[1em]
\includegraphics[clip,width=0.45\textwidth]{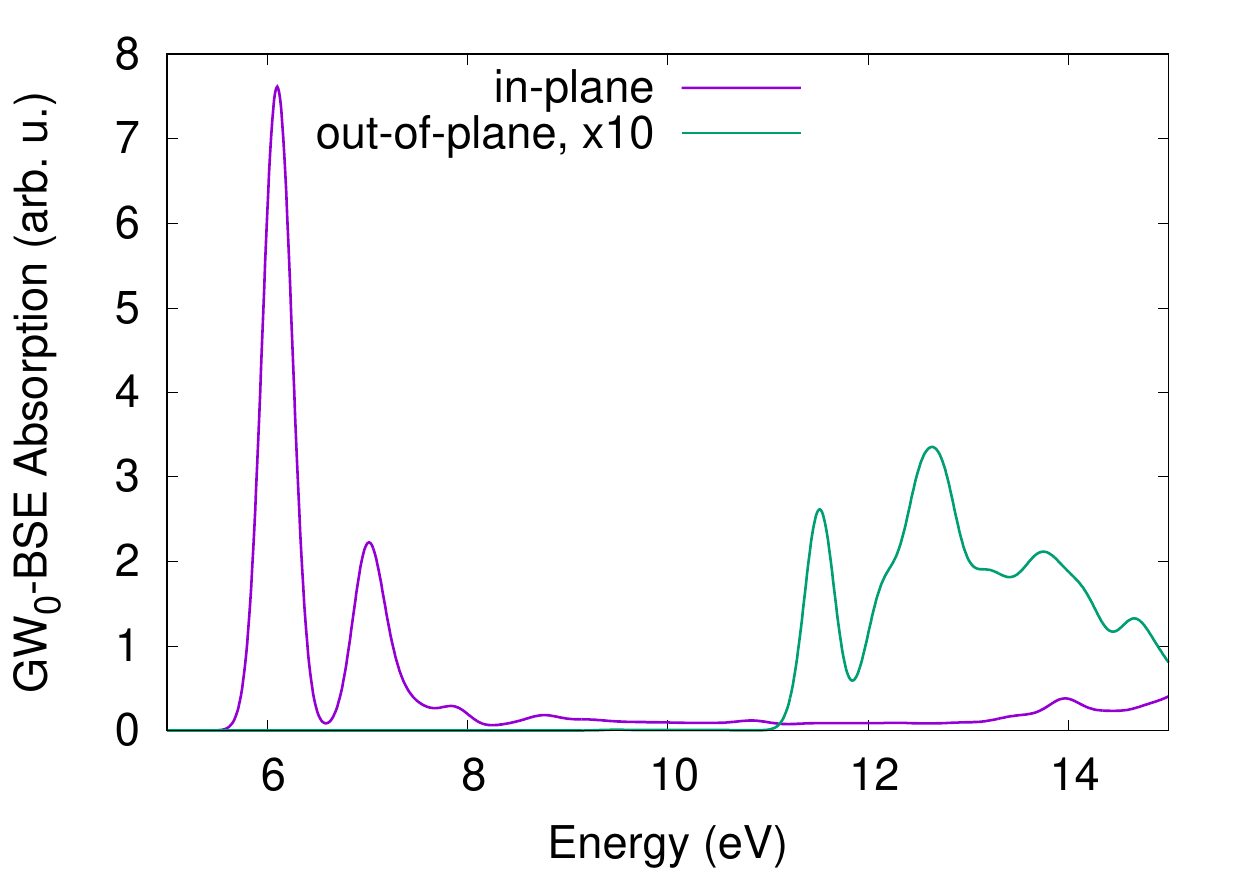}
\caption{(Color online) Electronic band structure of monolayer hBN,
  comparing DFT-PBE with $GW$ theory at the single-shot ($G_0W_0$) and
  partially self-consistent ($GW_0$) level (top panel). $GW_0$-BSE
  optical absorption spectrum of monolayer hBN for in- and out-of-plane
  polarization (bottom panel).
\label{fig:GW-BSE}}
\end{center}
\end{figure}

In Fig.\ \ref{fig:GW-BSE} we compare the electronic band structure
predicted by the two levels of $GW$ theory and DFT-PBE calculations
(top panel), and we plot the $GW_0$-BSE absorption spectrum (bottom
panel).  The exciton binding energy is extracted by comparing the BSE
optical absorption spectrum with its random-phase-approximation
counterpart, in which electron-hole interactions are neglected.  Our
single-shot $G_0W_0$ quasiparticle gaps are significantly smaller than
the SJ-DMC gaps, by about 1.4--2.5 eV\@.  The partially
self-consistent $GW_0$ quasiparticle gaps are somewhat larger, but are
still 1.1--2.2 eV smaller than the SJ-DMC quasiparticle gaps.  The
exciton binding energies obtained using first-principles $GW_0$-BSE
and SJ-DMC calculations are in reasonable agreement with the exciton
binding energies of $2.38$ eV ($\Gamma_{\rm v}\rightarrow K_{\rm c}$)
and $2.41$ eV ($K_{\rm v} \rightarrow K_{\rm c}$) obtained using an
effective-mass model \cite{Mostaani_2017} of an electron and a hole
interacting via the Keldysh interaction with the effective masses in
Table \ref{table:dft_eff_masses} and the $r_*$ parameter estimated in
Sec.\ \ref{sec:finite_size}.

DFT-LDA and DFT-PBE band-structure calculations are qualitatively
incorrect for monolayer hBN\@: they predict the gap to be direct
($K_{\rm v} \rightarrow K_{\rm c}$).  DFT-HSE06 and $GW$ calculations
show that the conduction-band energies at $K_{\rm c}$ and $M_{\rm c}$
are similar, but that the CBM lies at $\Gamma_{\rm c}$, in agreement
with SJ-DMC\@.

\begin{figure*}[!htbp]
\begin{center}
\includegraphics[width=0.47\textwidth,clip]{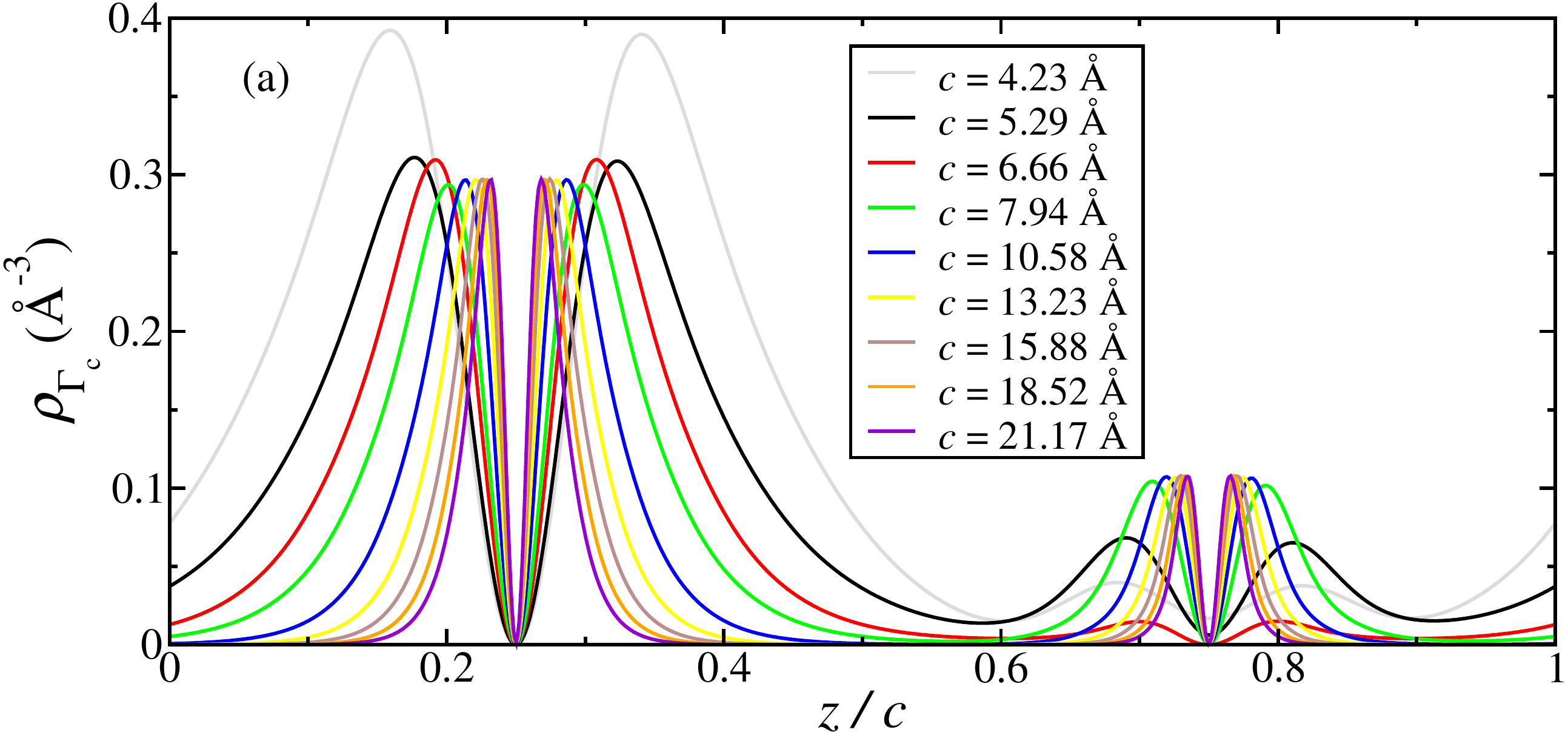} ~~
\includegraphics[width=0.47\textwidth,clip]{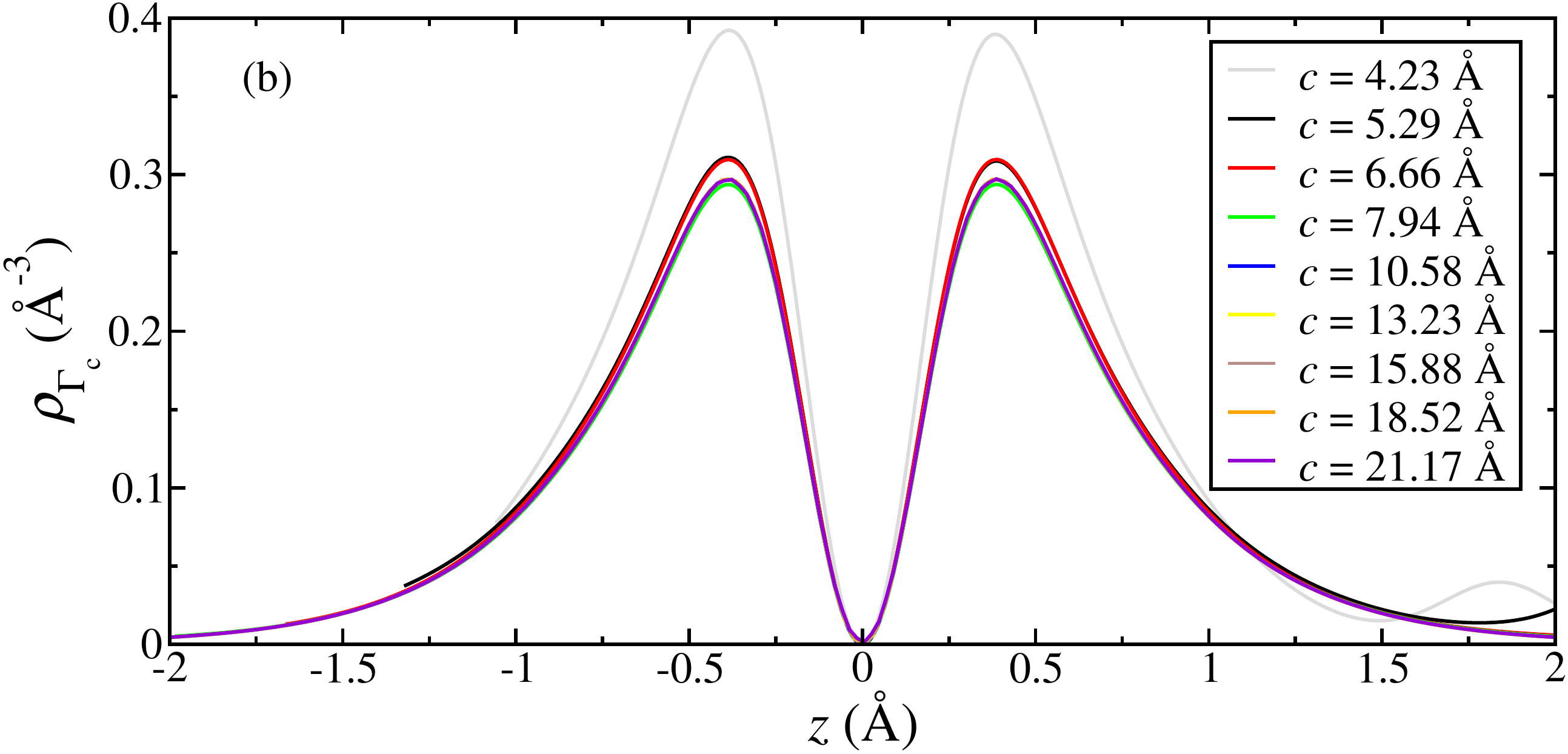}
\caption{(Color online) DFT-PBE charge density of the state at
  $\Gamma_{\rm c}$ as a function of lattice parameter $c$ for bulk
  hBN\@. $c=12.5878$ a.u.\ is the experimental lattice parameter.
  Panels (a) and (b) show the density against fractional and absolute
  $z$ coordinates, respectively.  The charge density is plotted along
  a straight line in the $z$ direction, passing through a boron atom
  at $z/c=0.25$ and a nitrogen atom at $z/c=0.75$.  At large $c$ the
  CBM at $\Gamma_{\rm c}$ is an arbitrary linear combination of the
  degenerate monolayer CBMs. \label{fig:crossover_gamma_1}}
\end{center}
\end{figure*}

\begin{figure}[!htbp]
\begin{center}
\includegraphics[width=0.45\textwidth,clip]{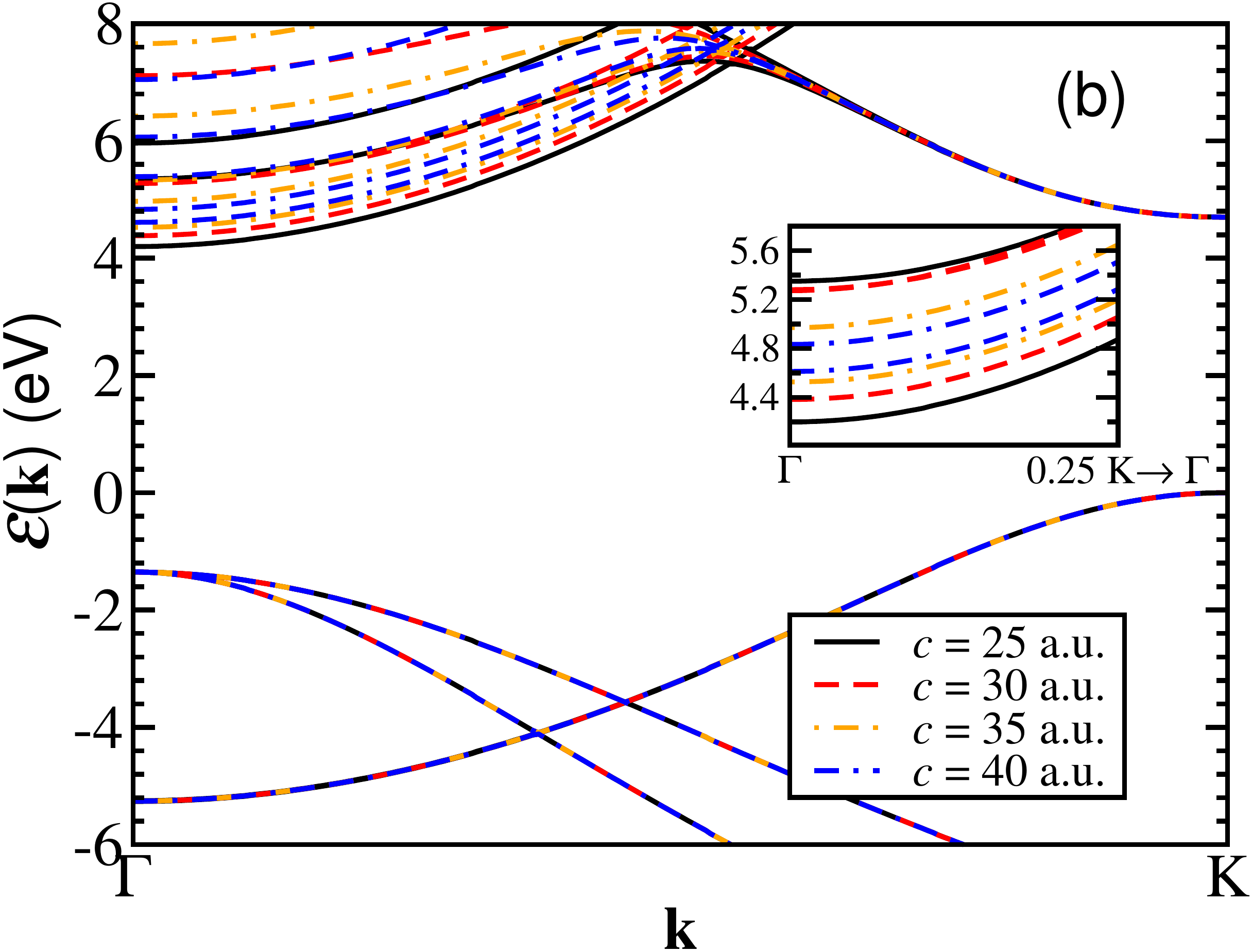}
\caption{(Color online) DFT-PBE bulk hBN band structure at three large
  values of the lattice parameter $c$. The inset to (b) displays a
  close-up of the two near-degenerate states at $\Gamma_{\rm
    c}$ \label{fig:crossover_gamma_2}}
\end{center}
\end{figure}

We find the gap of monolayer hBN to be indirect, with the CBM lying at
the $\Gamma_{\rm c}$ point, although recent experiments indicate a
direct gap at the ${\rm K}$ point of the Brillouin
zone \cite{Elias_2019}. Part of the reason for the discrepancy is that
those experiments studied hBN on a graphite substrate; however, the
delocalized nature of the (nearly free) CBM state at $\Gamma_{\rm c}$
may also have consequences for optical absorption
experiments. Electrons with small in-plane momentum experience the hBN
monolayer as an attractive $\delta$-function-like potential, always
supporting one bound state. This weakly bound state is potentially
sensitive to perturbations caused by substrates or other aspects of
the material environment. We have investigated the behavior of the
conduction band at $\Gamma_{\rm c}$ in bulk hBN as the out-of-plane
lattice parameter $c$ is increased, describing the crossover from bulk
to isolated monolayer.  In Fig.\ \ref{fig:crossover_gamma_1}, we plot
the normalized DFT-PBE charge density of the state at $\Gamma_{\rm c}$
along a line through the unit cell, moving through a boron atom at
$z/c=0.25$ and through a nitrogen atom at $z/c=0.75$. In
Fig.\ \ref{fig:crossover_gamma_2} we plot the DFT-PBE band structure
along the $\Gamma \rightarrow {\rm K}$ line.  While all other states
are well-converged with respect to $c$, including the state at ${\rm
  K}_{\rm c}$, the state at $\Gamma_{\rm c}$ remains relatively
sensitive to the particular choice of $c$. In the inset to
Fig.\ \ref{fig:crossover_gamma_2} the two lowest-lying conduction
states have been retained for clarity, and this sensitivity is made
very clear. The expected trend in the energy of the two
near-degenerate conduction states originating from each monolayer is
observed, and as $c$ increases, the energy splitting of these two
states reduces.

\subsubsection{Vibrational renormalization of the band structure}

Using a combination of the quadratic and stochastic approaches as
described in Sec.\ \ref{subsec:vib_method}, we obtain a vibrational
renormalization of the minimum band gap
$K_{\mathrm{v}}\to\Gamma_{\mathrm{c}}$ of monolayer hBN of $-0.56$ eV
at $0$ K\@. This zero temperature correction arises purely from quantum
zero-point motion, which has a strong effect in a system like hBN
containing light elements, and is similar in size to that calculated
for diamond \cite{Giustino_2010,Antonius_2014,Monserrat_2016}. Thermal
motion further renormalizes the band gap, resulting in a vibrational
correction of $-0.73$ eV at $300$ K\@.

Our results for the $K_{\mathrm{v}}\to K_{\mathrm{c}}$ gap show a
zero-point renormalization of the band gap of $-0.54$ eV, which
increases to $-0.73$ eV at $300$ K\@. The similar corrections for
the $K_{\mathrm{v}}\to\Gamma_{\mathrm{c}}$ and $K_{\mathrm{v}}\to
K_{\mathrm{c}}$ gaps suggest that vibrational corrections to the gap
are largely uniform across the Brillouin zone.

\subsubsection{Bulk hBN}

As a test of the accuracy of our methods, we have calculated the
quasiparticle and excitonic gaps of bulk hBN between various
high-symmetry points in the Brillouin zone with the QMC and $GW$
methods. Our QMC calculations are identical to those performed for the
monolayer, save for the use of the experimental geometry (lattice
parameters $a=2.504$ {\AA} and $c=6.6612$ {\AA}) \cite{Lynch_1966}, and
the use of the ``T-move'' scheme, which reduces pseudopotential
locality approximation
errors \cite{Casula_2006,Casula_2010,Drummond_2016}.

\begin{figure}[!htbp]
\begin{center}
\includegraphics[clip,width=0.45\textwidth]{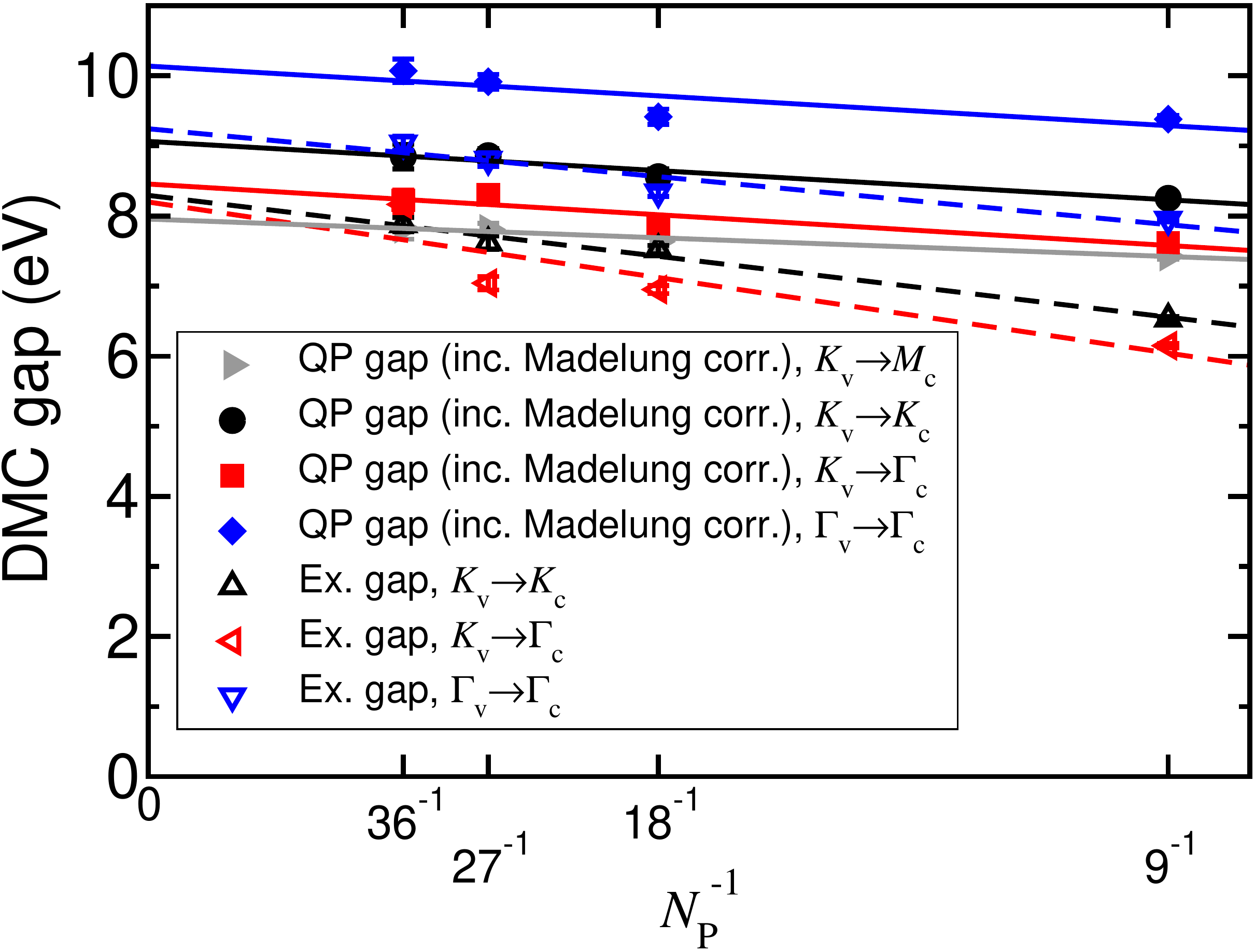}
  \caption{(Color online) SJ-DMC quasiparticle gaps $\Delta_{\rm qp}$ and
    excitonic gaps $\Delta_{\rm ex}$ of bulk hBN against $1/N_{\rm
      P}$, where $N_{\rm P}$ is the number of primitive cells in the
    supercell. The quasiparticle gaps include the Madelung correction
    given in Eq.\ (\ref{eq:qp_gap_corr}). The statistical error bars
    show the random error in the SJ-DMC gap in a particular supercell;
    the noise due to quasirandom finite-size effects clearly exceeds
    the noise due to the Monte Carlo calculation.}
\label{fig:bulk_fs}
\end{center}
\end{figure}

Our QMC results are given in Table \ref{table:bulk_gap_results}, with
error bars determined as discussed in Sec.\ \ref{sec:finite_size}. Our
raw gap data are plotted against system size in
Fig.\ \ref{fig:bulk_fs}. We find that quasirandom finite-size effects
are much more prominent in the bulk than in the monolayer. This could
be partially due to the lack of geometrical similarity of the
supercells studied, leading to nonsystematic behavior in the
charge-quadrupole finite-size effect.  Our $GW$ results for bulk hBN
are also shown in Table \ref{table:bulk_gap_results}. Here we find
that the quasiparticle gaps evaluated with SJ-DMC are somewhat larger
than those predicted by $GW$ calculations, just as they are in the
monolayer.

The SJ-DMC $K_{\rm v} \rightarrow K_{\rm c}$ exciton binding energy of bulk hBN,
which is corrected by the subtraction of the screened Madelung constant and
then extrapolated against $N^{-1}_{\rm P}$ to infinite system
size \cite{Hunt_2018}, is $0.8(1)$ eV\@.  This is consistent with the range of
$GW$-BSE values, and is significantly smaller than the monolayer exciton
binding energy, as one would expect.  The $K_{\rm v} \rightarrow \Gamma_{\rm
c}$ exciton binding is $0.3(5)$ eV, which is smaller than the statistical error
bars.

The SJ-DMC $K_{\rm v} \rightarrow M_{\rm c}$ quasiparticle gap is
7.96(9) eV\@.  The VBM in bulk hBN is near the $K$ point, while the
CBM is at or near the $M$ point \cite{Cassabois_2016}. Allowing for
our calculated zero-temperature vibrational correction in bulk hBN of
$-0.35$ eV (which increases to $-0.40$ eV at $300$ K), the SJ-DMC
quasiparticle gap appears to overestimate the experimental gap of
around 6 eV significantly.  As a probe of this discrepancy, we have
considered (in the $N_{\rm P}=9$ supercell) the effects of a backflow
transformation of the many-electron wave function. We have also
investigated our use of high-symmetry points ($K$ and $M$) in the
Brillouin zone rather than the true positions of the VBM and CBM at
the DFT-HSE06 level of theory.

We find that backflow lowers the DMC quasiparticle ($K_{\rm v}
\rightarrow$~CBM) gap of bulk hBN in the $N_{\rm P}=9$ supercell by
0.17(5) eV\@. By considering the exact VBM and CBM positions, we find
a further energy lowering of 0.02(6) eV, which is not statistically
significant.  Further, we have also considered explicit
re-optimization of backflow functions in anionic and cationic states
for the VBM~$\rightarrow$~CBM quasiparticle gap.  This has recently
been shown to lead to significant further lowering of SJB-DMC
quasiparticle energy gaps \cite{Hunt_2018}; however, in this case we
find that re-optimization of the backflow functions by minimizing the
VMC energy \textit{raises} the SJB-DMC gap by 0.08(3) eV as is also
found in the monolayer. Near-degeneracy of the bands at the $M$
point is a possible cause of both the unusual behavior of the DMC
energy in the presence of backflow and the overestimate of the
gap. Near-degeneracy can lead to multireference character and hence
significant fixed-node errors with a single-determinant wave function.

\begin{table*}[!htbp]
\begin{center}
\caption{Static-nucleus quasiparticle and excitonic gaps for bulk
  hBN, determined by different methods, compared with
  experimental results. Our DFT vibrational-renormalization
  calculations indicate that the static-nucleus gaps should be
  renormalized by $-0.40$ eV at 300 K\@. Where references are not
  given, the results are from the present work. An asterisk (*)
  denotes the SJ-DMC energy gap from $K_{\rm v} \rightarrow M_{\rm c}$.}
\begin{tabular}{lr@{}lr@{}lr@{}lr@{}lr@{}lr@{}lr@{}lr@{}lr@{}lr@{}l}
\hline \hline

 & \multicolumn{12}{c}{Quasiparticle gap $\Delta_{\rm qp}$ (eV)}
 & \multicolumn{8}{c}{Excitonic gap $\Delta_{\rm ex}$ (eV)} \\

\raisebox{1.5ex}[0pt]{Method}
    & \multicolumn{2}{c}{$\Gamma_{\rm v}\rightarrow \Gamma_{\rm c}$}
    & \multicolumn{2}{c}{$M_{\rm v}\rightarrow M_{\rm c}$}
    & \multicolumn{2}{c}{$K_{\rm v}\rightarrow \Gamma_{\rm c}$}
    & \multicolumn{2}{c}{$K_{\rm v}\rightarrow K_{\rm c}$}
    & \multicolumn{2}{c}{$M_{\rm v}\rightarrow \Gamma_{\rm c}$}
    & \multicolumn{2}{c}{VBM${}\rightarrow{}$CBM}
    & \multicolumn{2}{c}{$\Gamma_{\rm v}\rightarrow \Gamma_{\rm c}$}
    & \multicolumn{2}{c}{$K_{\rm v} \rightarrow \Gamma_{\rm c}$}
    & \multicolumn{2}{c}{$K_{\rm v}\rightarrow K_{\rm c}$}
    & \multicolumn{2}{c}{VBM${}\rightarrow{}$CBM} \\

\hline

DFT-LDA   &~~$6$&$.09$ & $4$&$.54$ & ~~~~$4$&$.93$ &~~$4$&$.84$ & $5$&$.28$ &
$4$&$.05$ & & & & & & & \\

DFT-PBE &$6$&$.65$ & $4$&$.76$ & $5$&$.42$ & $4$&$.94$ & $5$&$.78$ &
$4$&$.28$ & & & & & & & & \\

DFT-HSE06 &$8$&$.01$ & $6$&$.09$ & $6$&$.54$ & $6$&$.33$ & $6$&$.95$ &
$5$&$.55$ & & & & & & & & \\

$G_0W_0$ & $7$&$.3$ & ~~~~~$7$&$.0$ & & & $9$&$.7$ & ~~~~$6$&$.1$ &
~~~~~~~$5$&$.4$ & & & & & & & & \\

$GW_0$ & $7$&$.3$ & $7$&$.1$ & & & $9$&$.9$ & $6$&$.1$ & $5$&$.5$ & &
& & & & & & \\

$GW$ \cite{Arnaud_2006} & $8$&$.4$ & $6$&$.5$ & $6$&$.9$ & $6$&$.9$
& $7$&$.3$ & $5$&$.95$ & & & & & & & & \\

SJ-DMC & $10$&$.1(2)$ & & & $8$&$.5(2)$ & $9$&$.06(8)$ & & & $7$&$.96(9)^{*}$ & ~~$9$&$.2(2)$ &
~~~$8$&$.2(5)$ & ~~~~~$8$&$.3(1)$ & & \\

Exp.\ \cite{Watanabe_2004,Cassabois_2016} & & & & & & & & & & &
\multicolumn{2}{c}{5.971, 6.08} & & & & & & &
\multicolumn{2}{c}{5.822, 5.955} \\

\hline \hline
\end{tabular}
\label{table:bulk_gap_results}
\end{center}
\end{table*}

\section{Conclusions\label{sec:conclusions}}

We have performed DFT, $GW$, and SJ-DMC calculations to determine the
electronic structure of free-standing monolayer and bulk hBN\@.
Systematic finite-size errors in the SJ-DMC quasiparticle gaps fall off
as the reciprocal of the linear size of the simulation supercell, but
can be corrected by subtracting an appropriately screened Madelung
constant from the gap.  The remaining finite-size effects are
dominated by quasirandom oscillations as a function of system size,
arising from the fact that long-range oscillations in the
pair-correlation function are forced to be commensurate with the
supercell.  We find the SJ-DMC quasiparticle gap for the monolayer to be
indirect ($K_{\rm v} \rightarrow \Gamma_{\rm c}$) and of magnitude
8.8(3) eV\@, which is larger than the gap predicted by the $G_0W_0$,
$GW_0$, and $GW$ methods.  Our bulk SJ-DMC quasiparticle gaps are also
systematically larger than those predicted by $GW$
calculations \cite{Arnaud_2006}. Using DFT, we also find a sizeable
vibrational correction to the monolayer band gap of $-0.73$ eV at
$300$ K, and a vibrational correction of $-0.40$ eV to the bulk band
gap at $300$ K\@.

SJ-DMC shows that hBN exhibits large exciton binding energies of
$1.9(4)$ eV and $1.8(4)$ eV for the indirect ($K_{\rm v}\rightarrow
\Gamma_{\rm c}$) and direct ($K_{\rm v}\rightarrow K_{\rm c}$)
excitons in the monolayer. The latter binding energy is similar to the
value predicted by our $GW_0$-BSE calculation for the direct exciton
and compares well to previous $GW$-BSE
calculations \cite{Arnaud_2006,Wirtz_2006,Cunningham_2018}, as well as
the exciton binding energy obtained within the effective-mass
approximation with the Keldysh interaction between charge
carriers \cite{Mostaani_2017}. The predicted quasiparticle gaps of hBN
increase significantly as one goes from DFT with local functionals, to
DFT with hybrid functionals, to $G_0W_0$, to $GW_0$, to $GW$, to
SJ-DMC\@.

Comparing SJ-DMC gaps with experimental results for bulk hBN shows that
the SJ-DMC gaps are significantly too high, even when DFT-calculated
vibrational renormalizations are included; the overestimate is around
$1.5$ eV\@.  Several sources of error on a 0.1--0.3 eV energy scale
have been identified: uncertainties due to pseudopotentials, residual
finite-size errors after extrapolation of the noisy data to infinite
system size, and the need for a more complete treatment of dynamical
correlation effects through the use of backflow wave functions.  In
addition there are unquantified fixed-node errors arising from the use
of a single-determinant wave function.  Although we investigated very
small multideterminant wave functions for the monolayer, it is
possible that there could be significant uncanceled fixed-node errors
due to multireference character in some of the excited-state wave
functions.  The mismatch between the minima of the VMC and DMC
energies with respect to backflow functions gives some hint that this
might be the case.  A further possible cause of the disagreement with
experiment is the underestimate of the vibrational renormalization of
the gap.  Several materials exhibit vibrational corrections to the
band gap that are up to 50\% (although typically only 10--20\%) larger
when calculated using $GW$ theory or hybrid functionals rather than a
semilocal DFT functional \cite{Antonius_2014,Monserrat_2016}. In the
case of hBN, vibrational renormalizations of the band gap could
therefore be as large as $-1$ eV for the monolayer at $300$ K and
$-0.5$ eV for the bulk at $300$ K\@.

Static-nucleus self-consistent $GW$ calculations agree remarkably well
with the experimental quasiparticle gap of bulk hBN, but taking into
account vibrational effects we find that the $GW$ quasiparticle gap is
underestimated by about $0.4$ eV\@.  When vibrational effects are
included, single-shot $G_0W_0$ methods underestimate the experimental
gap by about $1$ eV\@.  Determining the electronic structure of hBN
from first principles with quantitative accuracy remains a challenging
problem.

\begin{acknowledgments}
We acknowledge financial support from the U.K.\ Engineering and
Physical Sciences Research Council (EPSRC) through a Science and
Innovation Award, the E.U.\ through the grant \textit{Concept
  Graphene}, the Royal Society, and Lancaster University through the
Early Career Small Grant Scheme. R.\ J.\ H.\ is fully funded by the
Graphene NOWNANO CDT (Grant No.\ EP/L01548X/1). B.\ M.\ acknowledges
support from the Winton Programme for the Physics of Sustainability,
and from Robinson College, Cambridge, and the Cambridge Philosophical
Society for a Henslow Research Fellowship. Computational resources
were provided by Lancaster University's High-End Computing
facility. VZ acknowledges the European Graphene Flagship Project and
the ARCHER National UK Supercomputer RAP Project e547. This work made
use of the facilities of N8 HPC provided and funded by the N8
consortium and EPSRC (Grant No.\ EP/K000225/1).  We acknowledge useful
discussions with V.\ I.\ Fal'ko, W.\ M.\ C.\ Foulkes, and S.\ Murphy.
\end{acknowledgments}

\end{document}